%% file: main.tex
\documentclass{IEEEtran4PSCC}
\ifCLASSINFOpdf
   \usepackage[pdftex]{graphicx}
\else
   \usepackage[dvips]{graphicx}
\fi
\usepackage[cmex10]{amsmath}
\usepackage{amssymb}

\usepackage{algorithm}
\usepackage{algpseudocode}
\usepackage{subcaption}
\usepackage{graphicx}

\usepackage{enumitem}

\usepackage{dblfloatfix}
\usepackage{subfiles}
\usepackage{tikz}
\usepackage{svg}
\usepackage{hyperref}

\usepackage{bbold}

\makeatletter
\let\old@ps@headings\ps@headings
\let\old@ps@IEEEtitlepagestyle\ps@IEEEtitlepagestyle
\def\psccfooter#1{%
    \def\ps@headings{%
        \old@ps@headings%
        \def\@oddfoot{\strut\hfill#1\hfill\strut}%
        \def\@evenfoot{\strut\hfill#1\hfill\strut}%
    }%
    \def\ps@IEEEtitlepagestyle{%
        \old@ps@IEEEtitlepagestyle%
        \def\@oddfoot{\strut\hfill#1\hfill\strut}%
        \def\@evenfoot{\strut\hfill#1\hfill\strut}%
    }%
    \ps@headings%
}
\makeatother

\psccfooter{%
}

\begin{document}
\title{Self-Supervised Graph Neural Networks\\for Optimal Substation Reconfiguration}

\author{\IEEEauthorblockN{Antoine Martinez\IEEEauthorrefmark{1},
Balthazar Donon\IEEEauthorrefmark{1},
Louis Wehenkel\IEEEauthorrefmark{2} and
Efthymios Karangelos\IEEEauthorrefmark{3}}
\IEEEauthorblockA{\IEEEauthorrefmark{1} Department of Research \& Development\\
RTE (Réseau de Transport d'Électricité), Paris, France}
\IEEEauthorblockA{\IEEEauthorrefmark{2} Department of Electrical Engineering \& Computer Science\\
Université de Liège, Liège, Belgium}
\IEEEauthorblockA{\IEEEauthorrefmark{3} School of Electrical and Electronic Engineering\\
University College Dublin, Dublin, Ireland}
}

\maketitle

\subfile{0_abstract}


\subfile{1_background}

\subfile{2_methodology}

\subfile{3_case_study}

\subfile{4_conclusion}

\bibliography{biblio}
\end{document}

%% file: 0_abstract.tex
\begin{abstract}

Changing the transmission system topology is an efficient and costless lever to reduce congestion or increase exchange capacities.
The problem of finding the optimal switch states within substations is called  \emph{Optimal Substation Reconfiguration} (OSR), and may be framed as a \emph{Mixed Integer Linear Program} (MILP).
Current state-of-the-art optimization techniques come with prohibitive computing times, making them impractical for real-time decision-making.
Meanwhile, deep learning offers a promising perspective with drastically smaller computing times, at the price of an expensive training phase and the absence of optimality guarantees.
In this work, we frame OSR as an \emph{Amortized Optimization} problem, where a \emph{Graph Neural Network} (GNN) model -- our data being graphs -- is trained in a self-supervised way to improve the objective function.
We apply our approach to the maximization of the exchange capacity between two areas of a small-scale 12-substations system.
Once trained, our GNN model improves the exchange capacity by 10.2\% on average compared to the all connected configuration, while a classical MILP solver reaches an average improvement of 15.2\% with orders-of-magnitude larger computing times.

\end{abstract}


%% file: 1_background.tex
\section{Introduction}

In light of increasing congestion costs from intensifying grid bottlenecks, the stakes for transmission grid topology optimization are higher than ever. 
The idea of rerouting power flows by opening or closing breakers and disconnector switches at negligible cost, rather than resorting to costly out-of-merit fossil-fueled generation, is advantageous both economically and environmentally. 
Unfortunately, the price to pay is the inherent computational complexity of the respective decision-making problem.
The academically-studied \emph{Optimal Transmission Switching} (OTS) problem, analytically stated as a \emph{Mixed Integer Programming Problem} (MILP),  already comes with an unmanageable computational burden \cite{hinneck} albeit involving (i) the simplified bus-branch model of the grid, which omits the internal configuration of substations and (ii) the linear \emph{Direct Current} (DC) approximation of the grid physics.
The practically-relevant \emph{Optimal Substation Reconfiguration} (OSR) variant, choosing how to operate switches determining connectivity within substations, further expands on this complexity \cite{MORSY}.

\subsection{Related Literature}
In recent years, the prospect of power grid topology optimization has been matched by the emerging computational opportunities offered by \emph{Machine Learning} (ML).
Yang and Oren \cite{yang} explored several supervised learning algorithms for quickly ranking candidate transmission lines to switch, in the style of the heuristic proposed by Fuller et al. \cite{fuller_fast_2012}.
This approach relies on a training dataset containing a large number of pre-generated network states with single-order transmission line switching actions and respective OPF solutions.
The limited ability of the tested algorithms to generalize over alternative grid topologies is a question that must be noted, especially with a view on the OSR problem.
The works in \cite{johnson2021knearestneighborheuristicrealtime,bugaje,pineda_learning-assisted_2024,Preuschoff} are alternative proposals that also rely on supervised learning algorithms and the availability of labeled data – specifically, power system operating snapshots paired with corresponding optimal topologies.
It is important to discuss here the critical assumption on the availability of labeled data.
In practice, historical data can at best reproduce the rule-based operation of the past power grid and may offer little insight into addressing the evolving challenges posed by the increasing penetration of renewable generation and the rise of the digital demand. 
Populating a suitable training dataset for OSR through additional computations would require analytically solving the ground-truth combinatorial problem en masse and at scale.
Such a task is computationally prohibitive and thus undermines the practicality of direct approaches to solve the OSR problem through supervised learning.

The alternative of relying on \emph{Reinforcement Learning} (RL) for power grid topology optimization has recently emerged as a distinct research direction.
The ability to learn by interacting with (any) simulator rather than relying on the availability of rich and informative training data is a potential advantage of RL in the context of  transmission grid topology optimization.
Rather than exhaustively describing the precise methodological features of state-of-the-art approaches (cf., \cite{lan,ramapuram,yoon_winning_2020,zhou,dorfer_power_2022}) it appears more relevant here to paint the bigger picture.
To date, there is no conclusive evidence that RL-based approaches for power grid topology optimization can outperform greedy-search algorithms and advanced rule-based algorithms integrating domain-specific knowledge 
\cite{lehna,sar_optimizing_2025}.
The curse of dimensionality, in other words the extremely large decision-space of alternative topologies, stands out once again as a notable challenge.
Although greedy reduction techniques have been very popularly explored (cf., \cite{lan,ramapuram,subramanian,jong_imitation_2024,jong_generalizable_2025,en15196920}) they typically impose a considerable computational burden and are inherently limited in their ability to capture the synergistic effects of multiple concurrent topological actions.
Interestingly, random sampling approaches have also been reported \cite{sar_optimizing_2025}, albeit with limited generalization potential and lacking any formal performance guarantee.

\subsection{Contributions}
This paper addresses the precise practical need to provide power system operators with decision-support tools for OSR that are compatible with real-time operation, even at the expense of not requiring strict optimality guarantees.
We pose the OSR problem in the \emph{Amortized Optimization} domain \cite{amortizedoptimization} and leverage self-supervised training to develop a model capable of learning to discover the most relevant topological actions through trial and error (single-step RL).

Our approach is enabled by recognizing power transmission grids as \emph{Hyper Heterogeneous Multi Graphs} (H2MGs) \cite{donon2022}, whose topological structure can largely vary from one operating condition to the other\footnote{See the open-source \href{https://huggingface.co/datasets/OpenSynth/D-GITT-RTE7000-2021}{RTE 7k dataset}.}.
Owing to such prominent topological variability, power grid data cannot be properly handled by most \emph{Deep Neural Network} (DNN) architectures which assume a constant number and ordering of objects that compose the grid. 
We overcome this inherent limitation of DNNs by using a \emph{Graph Neural Network} (GNN) architecture \cite{gnn_original, thomaskipfgnn}.
GNNs are a class of DNNs that are especially designed to handle graph data and robust to objects creation, deletion and permutation.
The particular GNN architecture introduced in \cite{donon2024} has already been shown to work efficiently on full-scale data without any heavy preprocessing.

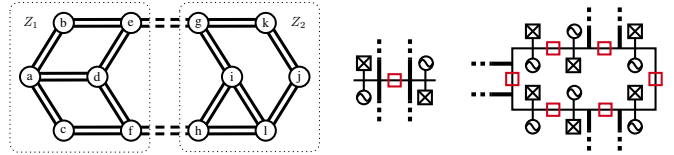
\begin{figure}[t!]
    \centering
    \begin{subfigure}[b]{0.48\linewidth}
        \centering
        \resizebox{\linewidth}{!}{\input{figures/case_study.tikz}}
        \caption{General structure of the case study substations and lines.}
        \label{fig:12_substations_system}
    \end{subfigure}
    \hfill
    \begin{subfigure}[b]{0.17\linewidth}
        \centering
        \resizebox{\linewidth}{!}{\input{figures/small_substation.tikz}}
        \caption{\textit{Type 1} substation.}
        \label{fig:small_substation}
    \end{subfigure}
    \hfill
    \begin{subfigure}[b]{0.31\linewidth}
        \centering
        \resizebox{\linewidth}{!}{\input{figures/large_substation.tikz}}
        \caption{\textit{Type 2}\\substation.}
        \label{fig:large_substation}
    \end{subfigure}    

    \caption{
        Twelve substations test case.
        Circles represent substations that are either of \textit{Type 1} (b) or of \textit{Type 2} (c), interconnected by double lines.
        The objective is to find which switches (in red) to open in order to maximize the transfer capacity from area $Z_1$ to area $Z_2$.
    }
    \label{fig:patricks_case}
\end{figure}

As a proof-of-concept study, this paper focuses on the problem of maximizing the exchange capacity between two neighboring areas, by changing the switch states.
{This is of special interest in the context of a border with multiple and spread-out connections (such as between Spain and France).}
More specifically, we consider the test case introduced in \cite{patrick} and illustrated in Figure \ref{fig:patricks_case}.

The major contributions of this work are:
\begin{itemize}[noitemsep,topsep=0pt]
    \item A methodology for the self-supervised training of a GNN to solve the OSR;
    \item An experimental validation of the methodology, along with a comparison with a classical optimization solver;
\end{itemize}

{The present paper falls in line with recent work \cite{authier2024physics} about the unsupervised training of a GNN to optimize switch states in a distribution system. The methodology is fairly different, as discrete variables were relaxed into continuous ones in \cite{authier2024physics}, and experiments were limited to a dataset two topological variants and to a smaller action space.}

\subsection{Paper Organization}

Section \ref{sec:methodology} defines the initial OSR problem as a MILP problem, transforms it into a continuous surrogate optimization problem and then into a statistical learning problem.
Section \ref{sec:case_study} details the dataset generation, the comparison baselines and the experimental protocol and results.
Finally, Section \ref{sec:conclusion} summarizes the methodological and experimental contributions of this work and proposes improvement areas required to get closer to an industrial application.

%% file: figures/case_study.tikz
\tikzset{every picture/.style={line width=0.75pt}} 

\begin{tikzpicture}[x=0.75pt,y=0.75pt,yscale=-1,xscale=1]

\draw [line width=2.25]    (22.36,94.27) -- (46.19,132.34) ;
\draw  [line width=1.5]  (43.72,30.5) .. controls (43.72,24.91) and (48.25,20.38) .. (53.84,20.38) .. controls (59.42,20.38) and (63.95,24.91) .. (63.95,30.5) .. controls (63.95,36.09) and (59.42,40.62) .. (53.84,40.62) .. controls (48.25,40.62) and (43.72,36.09) .. (43.72,30.5) -- cycle ;
\draw [line width=2.25]    (28.06,90.97) -- (51.74,128.44) ;
\draw [line width=2.25]    (89.81,94.27) -- (113.64,132.34) ;
\draw [line width=2.25]    (95.5,90.97) -- (119.18,128.44) ;
\draw [line width=2.25]    (224.69,94.27) -- (248.52,132.34) ;
\draw [line width=2.25]    (230.39,90.97) -- (254.06,128.44) ;
\draw [line width=2.25]    (258.41,40.32) -- (282.24,78.38) ;
\draw [line width=2.25]    (264.11,37.02) -- (287.79,74.49) ;
\draw [line width=2.25]    (190.97,40.32) -- (214.8,78.38) ;
\draw [line width=2.25]    (196.66,37.02) -- (220.34,74.49) ;
\draw [line width=2.25]    (29.56,81.03) -- (78.12,80.91) ;
\draw [line width=2.25]    (29.45,87.77) -- (78,87.66) ;
\draw [line width=2.25]    (63.28,27.07) -- (111.84,26.96) ;
\draw [line width=2.25]    (63.17,33.82) -- (111.72,33.7) ;
\draw [line width=2.25]  [dash pattern={on 6.75pt off 4.5pt}]  (130.72,27.07) -- (179.28,26.96) ;
\draw [line width=2.25]  [dash pattern={on 6.75pt off 4.5pt}]  (130.61,33.82) -- (179.17,33.7) ;
\draw [line width=2.25]    (198.16,27.07) -- (246.72,26.96) ;
\draw [line width=2.25]    (198.05,33.82) -- (246.61,33.7) ;
\draw [line width=2.25]    (63.28,134.98) -- (111.84,134.87) ;
\draw [line width=2.25]  [dash pattern={on 6.75pt off 4.5pt}]  (130.72,134.98) -- (179.28,134.87) ;
\draw [line width=2.25]    (198.16,134.98) -- (246.72,134.87) ;
\draw [line width=2.25]    (28.21,78.13) -- (51.72,40.23) ;
\draw [line width=2.25]    (22.62,74.47) -- (46.13,36.57) ;
\draw [line width=2.25]    (95.65,78.13) -- (119.16,40.23) ;
\draw [line width=2.25]    (90.06,74.47) -- (113.57,36.57) ;
\draw [line width=2.25]    (196.81,132.09) -- (220.32,94.18) ;
\draw [line width=2.25]    (191.23,128.43) -- (214.73,90.52) ;
\draw [line width=2.25]    (264.26,132.09) -- (287.76,94.18) ;
\draw [line width=2.25]    (258.67,128.43) -- (282.18,90.52) ;
\draw  [line width=1.5]  (111.16,30.5) .. controls (111.16,24.91) and (115.69,20.38) .. (121.28,20.38) .. controls (126.87,20.38) and (131.4,24.91) .. (131.4,30.5) .. controls (131.4,36.09) and (126.87,40.62) .. (121.28,40.62) .. controls (115.69,40.62) and (111.16,36.09) .. (111.16,30.5) -- cycle ;
\draw  [line width=1.5]  (77.44,84.45) .. controls (77.44,78.87) and (81.97,74.34) .. (87.56,74.34) .. controls (93.15,74.34) and (97.67,78.87) .. (97.67,84.45) .. controls (97.67,90.04) and (93.15,94.57) .. (87.56,94.57) .. controls (81.97,94.57) and (77.44,90.04) .. (77.44,84.45) -- cycle ;
\draw  [line width=1.5]  (10,84.45) .. controls (10,78.87) and (14.53,74.34) .. (20.12,74.34) .. controls (25.7,74.34) and (30.23,78.87) .. (30.23,84.45) .. controls (30.23,90.04) and (25.7,94.57) .. (20.12,94.57) .. controls (14.53,94.57) and (10,90.04) .. (10,84.45) -- cycle ;
\draw  [line width=1.5]  (43.72,138.41) .. controls (43.72,132.82) and (48.25,128.29) .. (53.84,128.29) .. controls (59.42,128.29) and (63.95,132.82) .. (63.95,138.41) .. controls (63.95,143.99) and (59.42,148.52) .. (53.84,148.52) .. controls (48.25,148.52) and (43.72,143.99) .. (43.72,138.41) -- cycle ;
\draw  [line width=1.5]  (111.16,138.41) .. controls (111.16,132.82) and (115.69,128.29) .. (121.28,128.29) .. controls (126.87,128.29) and (131.4,132.82) .. (131.4,138.41) .. controls (131.4,143.99) and (126.87,148.52) .. (121.28,148.52) .. controls (115.69,148.52) and (111.16,143.99) .. (111.16,138.41) -- cycle ;
\draw  [line width=1.5]  (178.6,30.5) .. controls (178.6,24.91) and (183.13,20.38) .. (188.72,20.38) .. controls (194.31,20.38) and (198.84,24.91) .. (198.84,30.5) .. controls (198.84,36.09) and (194.31,40.62) .. (188.72,40.62) .. controls (183.13,40.62) and (178.6,36.09) .. (178.6,30.5) -- cycle ;
\draw  [line width=1.5]  (246.05,30.5) .. controls (246.05,24.91) and (250.58,20.38) .. (256.16,20.38) .. controls (261.75,20.38) and (266.28,24.91) .. (266.28,30.5) .. controls (266.28,36.09) and (261.75,40.62) .. (256.16,40.62) .. controls (250.58,40.62) and (246.05,36.09) .. (246.05,30.5) -- cycle ;
\draw  [line width=1.5]  (279.77,84.45) .. controls (279.77,78.87) and (284.3,74.34) .. (289.88,74.34) .. controls (295.47,74.34) and (300,78.87) .. (300,84.45) .. controls (300,90.04) and (295.47,94.57) .. (289.88,94.57) .. controls (284.3,94.57) and (279.77,90.04) .. (279.77,84.45) -- cycle ;
\draw  [line width=1.5]  (212.33,84.45) .. controls (212.33,78.87) and (216.85,74.34) .. (222.44,74.34) .. controls (228.03,74.34) and (232.56,78.87) .. (232.56,84.45) .. controls (232.56,90.04) and (228.03,94.57) .. (222.44,94.57) .. controls (216.85,94.57) and (212.33,90.04) .. (212.33,84.45) -- cycle ;
\draw  [line width=1.5]  (246.05,138.41) .. controls (246.05,132.82) and (250.58,128.29) .. (256.16,128.29) .. controls (261.75,128.29) and (266.28,132.82) .. (266.28,138.41) .. controls (266.28,143.99) and (261.75,148.52) .. (256.16,148.52) .. controls (250.58,148.52) and (246.05,143.99) .. (246.05,138.41) -- cycle ;
\draw  [line width=1.5]  (178.6,138.41) .. controls (178.6,132.82) and (183.13,128.29) .. (188.72,128.29) .. controls (194.31,128.29) and (198.84,132.82) .. (198.84,138.41) .. controls (198.84,143.99) and (194.31,148.52) .. (188.72,148.52) .. controls (183.13,148.52) and (178.6,143.99) .. (178.6,138.41) -- cycle ;
\draw [line width=2.25]    (63.17,141.72) -- (111.72,141.61) ;
\draw [line width=2.25]  [dash pattern={on 6.75pt off 4.5pt}]  (130.61,141.72) -- (179.17,141.61) ;
\draw [line width=2.25]    (198.05,141.72) -- (246.61,141.61) ;
\draw  [draw opacity=0] (10,20) -- (300,20) -- (300,150) -- (10,150) -- cycle ;
\draw  [dash pattern={on 0.84pt off 2.51pt}] (0,18.53) .. controls (0,13.82) and (3.82,10) .. (8.53,10) -- (131.47,10) .. controls (136.18,10) and (140,13.82) .. (140,18.53) -- (140,151.47) .. controls (140,156.18) and (136.18,160) .. (131.47,160) -- (8.53,160) .. controls (3.82,160) and (0,156.18) .. (0,151.47) -- cycle ;
\draw  [dash pattern={on 0.84pt off 2.51pt}] (170,18.53) .. controls (170,13.82) and (173.82,10) .. (178.53,10) -- (301.47,10) .. controls (306.18,10) and (310,13.82) .. (310,18.53) -- (310,150.93) .. controls (310,155.64) and (306.18,159.45) .. (301.47,159.45) -- (178.53,159.45) .. controls (173.82,159.45) and (170,155.64) .. (170,150.93) -- cycle ;

\draw (19.84,84.45) node  [font=\normalsize] [align=left] {\begin{minipage}[lt]{8.67pt}\setlength\topsep{0pt}
\begin{center}
a
\end{center}

\end{minipage}};
\draw (53.56,30.5) node  [font=\normalsize] [align=left] {\begin{minipage}[lt]{8.67pt}\setlength\topsep{0pt}
\begin{center}
b
\end{center}

\end{minipage}};
\draw (53.56,138.41) node  [font=\normalsize] [align=left] {\begin{minipage}[lt]{8.67pt}\setlength\topsep{0pt}
\begin{center}
c
\end{center}

\end{minipage}};
\draw (87.28,84.45) node  [font=\normalsize] [align=left] {\begin{minipage}[lt]{8.67pt}\setlength\topsep{0pt}
\begin{center}
d
\end{center}

\end{minipage}};
\draw (121,30.5) node  [font=\normalsize] [align=left] {\begin{minipage}[lt]{8.67pt}\setlength\topsep{0pt}
\begin{center}
e
\end{center}

\end{minipage}};
\draw (121,138.41) node  [font=\normalsize] [align=left] {\begin{minipage}[lt]{8.67pt}\setlength\topsep{0pt}
\begin{center}
f
\end{center}

\end{minipage}};
\draw (188.72,30.5) node  [font=\normalsize] [align=left] {\begin{minipage}[lt]{8.67pt}\setlength\topsep{0pt}
\begin{center}
g
\end{center}

\end{minipage}};
\draw (256.16,30.5) node  [font=\normalsize] [align=left] {\begin{minipage}[lt]{8.67pt}\setlength\topsep{0pt}
\begin{center}
k
\end{center}

\end{minipage}};
\draw (222.44,84.45) node  [font=\normalsize] [align=left] {\begin{minipage}[lt]{8.67pt}\setlength\topsep{0pt}
\begin{center}
i
\end{center}

\end{minipage}};
\draw (289.88,84.45) node  [font=\normalsize] [align=left] {\begin{minipage}[lt]{8.67pt}\setlength\topsep{0pt}
\begin{center}
j
\end{center}

\end{minipage}};
\draw (188.72,138.41) node  [font=\normalsize] [align=left] {\begin{minipage}[lt]{8.67pt}\setlength\topsep{0pt}
\begin{center}
h
\end{center}

\end{minipage}};
\draw (256.16,138.41) node  [font=\normalsize] [align=left] {\begin{minipage}[lt]{8.67pt}\setlength\topsep{0pt}
\begin{center}
l
\end{center}

\end{minipage}};
\draw (12,23.4) node [anchor=north west][inner sep=0.75pt]    {$Z_{1}$};
\draw (298,23.4) node [anchor=north east] [inner sep=0.75pt]    {$Z_{2}$};

\end{tikzpicture}

%% file: figures/small_substation.tikz
\tikzset{every picture/.style={line width=0.75pt}} 

\begin{tikzpicture}[x=0.75pt,y=0.75pt,yscale=-1,xscale=1]

\draw  [line width=1.5]  (523.33,56.67) -- (536.67,56.67) -- (536.67,70) -- (523.33,70) -- cycle ;
\draw [line width=1.5]    (523.33,56.67) -- (536.67,70) ;
\draw [line width=1.5]    (523.33,70) -- (536.67,56.67) ;

\draw [line width=1.5]    (530,70) -- (530,90) ;
\draw  [color={rgb, 255:red, 0; green, 0; blue, 0 }  ,draw opacity=1 ][line width=1.5]  (523.33,96.67) .. controls (524.42,94.96) and (525.46,93.33) .. (526.67,93.33) .. controls (527.87,93.33) and (528.91,94.96) .. (530,96.67) .. controls (531.09,98.37) and (532.13,100) .. (533.33,100) .. controls (534.54,100) and (535.58,98.37) .. (536.67,96.67) ;
\draw  [color={rgb, 255:red, 0; green, 0; blue, 0 }  ,draw opacity=1 ][line width=1.5]  (523.33,96.67) .. controls (523.33,92.98) and (526.32,90) .. (530,90) .. controls (533.68,90) and (536.67,92.98) .. (536.67,96.67) .. controls (536.67,100.35) and (533.68,103.33) .. (530,103.33) .. controls (526.32,103.33) and (523.33,100.35) .. (523.33,96.67) -- cycle ;

\draw [line width=3]    (545,60) -- (545,80) ;
\draw [line width=3]    (575,60) -- (575,80) ;
\draw [line width=3]  [dash pattern={on 3.38pt off 3.27pt}]  (575,40) -- (575,60) ;
\draw [line width=3]  [dash pattern={on 3.38pt off 3.27pt}]  (545,40) -- (545,60) ;
\draw [line width=3]    (545,80) -- (545,100) ;
\draw [line width=3]    (575,80) -- (575,100) ;
\draw [line width=3]  [dash pattern={on 3.38pt off 3.27pt}]  (545,120) -- (545,100) ;
\draw [line width=3]  [dash pattern={on 3.38pt off 3.27pt}]  (575,120) -- (575,100) ;
\draw  [color={rgb, 255:red, 0; green, 0; blue, 0 }  ,draw opacity=1 ][line width=1.5]  (584,63.33) .. controls (585.09,61.63) and (586.13,60) .. (587.33,60) .. controls (588.54,60) and (589.58,61.63) .. (590.67,63.33) .. controls (591.75,65.04) and (592.79,66.67) .. (594,66.67) .. controls (595.16,66.67) and (596.16,65.17) .. (597.2,63.54) ;
\draw  [color={rgb, 255:red, 0; green, 0; blue, 0 }  ,draw opacity=1 ][line width=1.5]  (584,63.33) .. controls (584,59.65) and (586.98,56.67) .. (590.67,56.67) .. controls (594.35,56.67) and (597.33,59.65) .. (597.33,63.33) .. controls (597.33,67.02) and (594.35,70) .. (590.67,70) .. controls (586.98,70) and (584,67.02) .. (584,63.33) -- cycle ;

\draw [line width=1.5]    (590,70) -- (590,90) ;
\draw  [line width=1.5]  (583,90) -- (596.33,90) -- (596.33,103.33) -- (583,103.33) -- cycle ;
\draw [line width=1.5]    (583,90) -- (596.33,103.33) ;
\draw [line width=1.5]    (583,103.33) -- (596.33,90) ;

\draw [line width=1.5]    (520,80) -- (600,80) ;
\draw  [color={rgb, 255:red, 208; green, 2; blue, 27 }  ,draw opacity=1 ][line width=1.5]  (554,74) -- (566,74) -- (566,86) -- (554,86) -- cycle ;
\draw  [draw opacity=0] (510,10) -- (610,10) -- (610,150) -- (510,150) -- cycle ;

\end{tikzpicture}

%% file: figures/large_substation.tikz
\tikzset{every picture/.style={line width=0.75pt}} 

\begin{tikzpicture}[x=0.75pt,y=0.75pt,yscale=-1,xscale=1]

\draw [line width=3]    (360,65) -- (340,65) ;
\draw [line width=3]    (360,95) -- (340,95) ;
\draw  [line width=1.5]  (360,50) -- (500,50) -- (500,110) -- (360,110) -- cycle ;
\draw [line width=3]    (435,30) -- (435,50) ;
\draw [line width=3]    (465,30) -- (465,50) ;
\draw  [line width=1.5]  (473.33,86.67) -- (486.67,86.67) -- (486.67,100) -- (473.33,100) -- cycle ;
\draw [line width=1.5]    (473.33,86.67) -- (486.67,100) ;
\draw [line width=1.5]    (473.33,100) -- (486.67,86.67) ;

\draw [line width=1.5]    (480,100) -- (480,120) ;
\draw  [color={rgb, 255:red, 0; green, 0; blue, 0 }  ,draw opacity=1 ][line width=1.5]  (473.33,126.67) .. controls (474.42,124.96) and (475.46,123.33) .. (476.67,123.33) .. controls (477.87,123.33) and (478.91,124.96) .. (480,126.67) .. controls (481.09,128.37) and (482.13,130) .. (483.33,130) .. controls (484.54,130) and (485.58,128.37) .. (486.67,126.67) .. controls (486.67,126.67) and (486.67,126.67) .. (486.67,126.67) ;
\draw  [color={rgb, 255:red, 0; green, 0; blue, 0 }  ,draw opacity=1 ][line width=1.5]  (473.33,126.67) .. controls (473.33,122.98) and (476.32,120) .. (480,120) .. controls (483.68,120) and (486.67,122.98) .. (486.67,126.67) .. controls (486.67,130.35) and (483.68,133.33) .. (480,133.33) .. controls (476.32,133.33) and (473.33,130.35) .. (473.33,126.67) -- cycle ;

\draw [line width=3]    (435,110) -- (435,130) ;
\draw [line width=3]    (465,110) -- (465,130) ;
\draw  [color={rgb, 255:red, 208; green, 2; blue, 27 }  ,draw opacity=1 ][line width=1.5]  (394,104) -- (406,104) -- (406,116) -- (394,116) -- cycle ;
\draw  [color={rgb, 255:red, 208; green, 2; blue, 27 }  ,draw opacity=1 ][line width=1.5]  (354,74) -- (366,74) -- (366,86) -- (354,86) -- cycle ;
\draw  [color={rgb, 255:red, 208; green, 2; blue, 27 }  ,draw opacity=1 ][line width=1.5]  (445,104) -- (457,104) -- (457,116) -- (445,116) -- cycle ;
\draw  [color={rgb, 255:red, 208; green, 2; blue, 27 }  ,draw opacity=1 ][line width=1.5]  (494,74) -- (506,74) -- (506,86) -- (494,86) -- cycle ;
\draw  [color={rgb, 255:red, 208; green, 2; blue, 27 }  ,draw opacity=1 ][line width=1.5]  (444,44) -- (456,44) -- (456,56) -- (444,56) -- cycle ;
\draw  [color={rgb, 255:red, 208; green, 2; blue, 27 }  ,draw opacity=1 ][line width=1.5]  (394,44) -- (406,44) -- (406,56) -- (394,56) -- cycle ;
\draw [line width=3]  [dash pattern={on 3.38pt off 3.27pt}]  (320,65) -- (340,65) ;
\draw [line width=3]  [dash pattern={on 3.38pt off 3.27pt}]  (320,95) -- (340,95) ;
\draw [line width=3]  [dash pattern={on 3.38pt off 3.27pt}]  (465,10) -- (465,30) ;
\draw [line width=3]  [dash pattern={on 3.38pt off 3.27pt}]  (435,10) -- (435,30) ;
\draw [line width=3]  [dash pattern={on 3.38pt off 3.27pt}]  (435,150) -- (435,130) ;
\draw [line width=3]  [dash pattern={on 3.38pt off 3.27pt}]  (465,150) -- (465,130) ;
\draw  [color={rgb, 255:red, 0; green, 0; blue, 0 }  ,draw opacity=1 ][line width=1.5]  (414,93.33) .. controls (415.09,91.63) and (416.13,90) .. (417.33,90) .. controls (418.54,90) and (419.58,91.63) .. (420.67,93.33) .. controls (421.75,95.04) and (422.79,96.67) .. (424,96.67) .. controls (425.16,96.67) and (426.16,95.17) .. (427.2,93.54) ;
\draw  [color={rgb, 255:red, 0; green, 0; blue, 0 }  ,draw opacity=1 ][line width=1.5]  (414,93.33) .. controls (414,89.65) and (416.98,86.67) .. (420.67,86.67) .. controls (424.35,86.67) and (427.33,89.65) .. (427.33,93.33) .. controls (427.33,97.02) and (424.35,100) .. (420.67,100) .. controls (416.98,100) and (414,97.02) .. (414,93.33) -- cycle ;

\draw [line width=1.5]    (420,100) -- (420,120) ;
\draw  [line width=1.5]  (413,120) -- (426.33,120) -- (426.33,133.33) -- (413,133.33) -- cycle ;
\draw [line width=1.5]    (413,120) -- (426.33,133.33) ;
\draw [line width=1.5]    (413,133.33) -- (426.33,120) ;

\draw  [line width=1.5]  (373.33,86.67) -- (386.67,86.67) -- (386.67,100) -- (373.33,100) -- cycle ;
\draw [line width=1.5]    (373.33,86.67) -- (386.67,100) ;
\draw [line width=1.5]    (373.33,100) -- (386.67,86.67) ;

\draw [line width=1.5]    (380,100) -- (380,120) ;
\draw  [color={rgb, 255:red, 0; green, 0; blue, 0 }  ,draw opacity=1 ][line width=1.5]  (373.33,126.67) .. controls (374.42,124.96) and (375.46,123.33) .. (376.67,123.33) .. controls (377.87,123.33) and (378.91,124.96) .. (380,126.67) .. controls (381.09,128.37) and (382.13,130) .. (383.33,130) .. controls (384.54,130) and (385.58,128.37) .. (386.67,126.67) .. controls (386.67,126.67) and (386.67,126.67) .. (386.67,126.67) ;
\draw  [color={rgb, 255:red, 0; green, 0; blue, 0 }  ,draw opacity=1 ][line width=1.5]  (373.33,126.67) .. controls (373.33,122.98) and (376.32,120) .. (380,120) .. controls (383.68,120) and (386.67,122.98) .. (386.67,126.67) .. controls (386.67,130.35) and (383.68,133.33) .. (380,133.33) .. controls (376.32,133.33) and (373.33,130.35) .. (373.33,126.67) -- cycle ;

\draw  [line width=1.5]  (373.33,26.67) -- (386.67,26.67) -- (386.67,40) -- (373.33,40) -- cycle ;
\draw [line width=1.5]    (373.33,26.67) -- (386.67,40) ;
\draw [line width=1.5]    (373.33,40) -- (386.67,26.67) ;

\draw [line width=1.5]    (380,40) -- (380,60) ;
\draw  [color={rgb, 255:red, 0; green, 0; blue, 0 }  ,draw opacity=1 ][line width=1.5]  (373.33,66.67) .. controls (374.42,64.96) and (375.46,63.33) .. (376.67,63.33) .. controls (377.87,63.33) and (378.91,64.96) .. (380,66.67) .. controls (381.09,68.37) and (382.13,70) .. (383.33,70) .. controls (384.54,70) and (385.58,68.37) .. (386.67,66.67) .. controls (386.67,66.67) and (386.67,66.67) .. (386.67,66.67) ;
\draw  [color={rgb, 255:red, 0; green, 0; blue, 0 }  ,draw opacity=1 ][line width=1.5]  (373.33,66.67) .. controls (373.33,62.98) and (376.32,60) .. (380,60) .. controls (383.68,60) and (386.67,62.98) .. (386.67,66.67) .. controls (386.67,70.35) and (383.68,73.33) .. (380,73.33) .. controls (376.32,73.33) and (373.33,70.35) .. (373.33,66.67) -- cycle ;

\draw  [line width=1.5]  (473.33,26.67) -- (486.67,26.67) -- (486.67,40) -- (473.33,40) -- cycle ;
\draw [line width=1.5]    (473.33,26.67) -- (486.67,40) ;
\draw [line width=1.5]    (473.33,40) -- (486.67,26.67) ;

\draw [line width=1.5]    (480,40) -- (480,60) ;
\draw  [color={rgb, 255:red, 0; green, 0; blue, 0 }  ,draw opacity=1 ][line width=1.5]  (473.33,66.67) .. controls (474.42,64.96) and (475.46,63.33) .. (476.67,63.33) .. controls (477.87,63.33) and (478.91,64.96) .. (480,66.67) .. controls (481.09,68.37) and (482.13,70) .. (483.33,70) .. controls (484.54,70) and (485.58,68.37) .. (486.67,66.67) .. controls (486.67,66.67) and (486.67,66.67) .. (486.67,66.67) ;
\draw  [color={rgb, 255:red, 0; green, 0; blue, 0 }  ,draw opacity=1 ][line width=1.5]  (473.33,66.67) .. controls (473.33,62.98) and (476.32,60) .. (480,60) .. controls (483.68,60) and (486.67,62.98) .. (486.67,66.67) .. controls (486.67,70.35) and (483.68,73.33) .. (480,73.33) .. controls (476.32,73.33) and (473.33,70.35) .. (473.33,66.67) -- cycle ;

\draw  [color={rgb, 255:red, 0; green, 0; blue, 0 }  ,draw opacity=1 ][line width=1.5]  (414,33.33) .. controls (415.09,31.63) and (416.13,30) .. (417.33,30) .. controls (418.54,30) and (419.58,31.63) .. (420.67,33.33) .. controls (421.75,35.04) and (422.79,36.67) .. (424,36.67) .. controls (425.16,36.67) and (426.16,35.17) .. (427.2,33.54) ;
\draw  [color={rgb, 255:red, 0; green, 0; blue, 0 }  ,draw opacity=1 ][line width=1.5]  (414,33.33) .. controls (414,29.65) and (416.98,26.67) .. (420.67,26.67) .. controls (424.35,26.67) and (427.33,29.65) .. (427.33,33.33) .. controls (427.33,37.02) and (424.35,40) .. (420.67,40) .. controls (416.98,40) and (414,37.02) .. (414,33.33) -- cycle ;

\draw [line width=1.5]    (420,40) -- (420,60) ;
\draw  [line width=1.5]  (413,60) -- (426.33,60) -- (426.33,73.33) -- (413,73.33) -- cycle ;
\draw [line width=1.5]    (413,60) -- (426.33,73.33) ;
\draw [line width=1.5]    (413,73.33) -- (426.33,60) ;

\draw  [draw opacity=0] (320,10) -- (510,10) -- (510,150) -- (320,150) -- cycle ;

\end{tikzpicture}

%% file: 2_methodology.tex
\section{Methodology}
\label{sec:methodology}

Let us describe how a GNN can be trained in an unsupervised fashion to tackle the OSR problem.

    \subsection{Initial Optimization Problem Description}

    First, let us frame the OSR as a MILP using the mathematical framework of H2MGs, which is key to successfully applying our overall methodology.

        \paragraph{Context Variable $x$}
        Let $x \in  \mathcal{X}$ be a power grid operating condition, which encompasses both its topological \textit{structure} and its \textit{features}.
        It is framed as a H2MG, made of:
        \begin{itemize}[noitemsep,topsep=0pt]
            \item A set of generators $\mathcal{E}_x^\text{gen}$;
            \item A set of loads $\mathcal{E}_x^\text{load}$;
            \item A set of switches $\mathcal{E}_x^\text{switch}$;
            \item A set of transmission lines $\mathcal{E}_x^\text{line}$.
        \end{itemize}
        These hyper-edges are connected to a series of addresses $\mathcal{A}_x \subset \mathbb{N}$ through class-specific ports (\textit{i.e.} mappings that associate each hyper-edge $e$ with a unique address $a\in \mathcal{A}_x$).
        Generators and loads have a single port denoted by $o$, while switches and transmission lines have two of them, namely $o^f$ (from) and $o^t$ (to).
        Let $\mathcal{O}^c$ denote the ports of a class $c\in \mathcal{C}:=\{\text{gen}, \text{load}, \text{switch}, \text{line}\}$. 
        
        Under the DC approximation hypotheses, a generator or a load $e \in \mathcal{E}_x^\text{gen} \cup \mathcal{E}_x^\text{load}$ is defined by:
        \begin{itemize}[noitemsep,topsep=0pt]
            \item $P_e$ its active power;
            \item $\mathbb{1}_e^{Z_1}$ (resp. $\mathbb{1}_e^{Z_2}$) a binary variable equal to $1$ if it belongs to area $Z_1$ (resp. $Z_2$) and $0$ otherwise.
        \end{itemize}
        A transmission line $e \in \mathcal{E}_x^\text{line}$ is defined by:
        \begin{itemize}[noitemsep,topsep=0pt]
            \item $\overline{F}_e$ its thermal limit;
            \item $X_e$ its reactance;
            \item $S_e$ a discrete variable equal to $+1$ if it points from area $Z_1$ to area $Z_2$, to $-1$ if it points from $Z_2$ to $Z_1$, and to $0$ if it is not on the boundary between $Z_1$ and $Z_2$.
        \end{itemize}

        \paragraph{Decision Variable $y$}
        For a given $x\in \mathcal{X}$, let us consider the variable $y \in \{0,1\}^{\mathcal{E}_x^\text{switch}}$, that associates each switch in $x$ with a binary state ($0$ for \textit{open} and $1$ for \textit{closed}).
        As illustrated in Figure \ref{fig:decision_variable}, this variable defines the actual topology of the grid.

        \paragraph{Objective Function $f$}
        For a given context $x$ and decision variable $y$, let us define the exchange capacity  as the flow from $Z_1$ to $Z_2$ obtained by homothetically increasing production in $Z_1$ and consumption in $Z_2$ by a factor $\lambda>0$ until a thermal limit is reached.
        
        The real-valued objective function $f(y;x)$ to be minimized is expressed by a Linear Program (LP) detailed in \cite{patrick}, {$-f(y;x)$ being the exchange capacity for a given topology $y$.}
        We formulate it using our H2MG formalism as follows,
        \begin{subequations}
        \label{eq:lp}
        \begin{align}
            f(y;x) &:= \underset{\substack{\lambda, (\vartheta_a)_{a \in \mathcal{A}_x}, \\(F_{e})_{e \in \mathcal{E}_x^{\text{switch}}}}}{\min} 
            \quad \underset{e\in \mathcal{E}_x^{\text{line}}}{\sum} S_{e} \frac{\vartheta_{o^t(e)}-\vartheta_{o^f(e)}}{X_{e}}  \label{eq:lp_objective}\\
            \text{s.t.} \quad   &\forall e \in \mathcal{E}_x^\text{switch}, \nonumber \\
                                &\quad \quad \vert \vartheta_{o^t(e)}-\vartheta_{o^f(e)} \vert \leq M (1-y_{e}) \label{eq:lp_breaker_closed}\\
                                &\quad \quad \vert F_{e} \vert \leq M y_{e} \label{eq:lp_breaker_open}\\
                                &\forall e \in \mathcal{E}_x^\text{line}, \nonumber \\
                                &\quad \quad \left| \frac{\vartheta_{o^t(e)}-\vartheta_{o^f(e)}}{X_{e}} \right| \leq  \overline{F}_{e} \label{eq:lp_line_flow}\\
                                &\forall a \in \mathcal{A}_x, \nonumber \\
                                &\quad \quad -\sum_{e \in \mathcal{E}_x^\text{gen} | o(e)=a} P_e (\lambda \mathbb{1}_e^{Z_1} + \mathbb{1}_e^{Z_2}) \nonumber \\
                                &\quad \quad +\sum_{e \in \mathcal{E}_x^\text{load} | o(e)=a} P_e (\mathbb{1}_e^{Z_1} + \mu \mathbb{1}_e^{Z_2}) \nonumber \\
                                &\quad \quad +\sum_{e\in \mathcal{E}_x^\text{switch} | o^f(e)=a} F_e  -\sum_{e\in \mathcal{E}_x^\text{switch} | o^t(e)=a} F_e \nonumber \\
                                &\quad \quad +\sum_{e\in \mathcal{E}_x^\text{line} | o^f(e)=a} \frac{\vartheta_{o^f(e)}-\vartheta_{o^t(e)}}{X_{e}} \nonumber \\
                                &\quad \quad -\sum_{e\in \mathcal{E}_x^\text{line} | o^t(e)=a} \frac{\vartheta_{o^f(e)}-\vartheta_{o^t(e)}}{X_{e}} = 0 \label{eq:lp_address_balance} \\
                                &\mu = \frac{ \sum\limits_{e \in \mathcal{E}_x^\text{gen}} P_e (\lambda \mathbb{1}_e^{Z_1} + \mathbb{1}_e^{Z_2}) - \underset{e\in \mathcal{E}_x^\text{load}}{\sum} P_e \mathbb{1}_e^{Z_1}  }{\underset{e \in \mathcal{E}_x^\text{load}}{\sum} P_e \mathbb{1}_e^{Z_2}} \label{eq:lp_coeff} \\
                                &\vartheta_1 = 0 \label{eq:lp_address_slack}
        \end{align}
        \end{subequations}
        where the unknowns are the multiplicative factor $\lambda$, the phase angles $(\vartheta_a)_{a \in \mathcal{A}_x}$ of the addresses and the power flows $(F_{e})_{e \in \mathcal{E}_x^\text{switch}}$ through the switches.
        The objective to minimize \eqref{eq:lp_objective} is the total active power flowing from $Z_2$ to $Z_1$ (which amounts to maximizing the flow from $Z_1$ to $Z_2$).
        Inequality \eqref{eq:lp_breaker_closed} imposes that phase angles at each end of a switch are equal if the latter is closed, while inequality \eqref{eq:lp_breaker_open} forces the power flowing through it to be null if it is open. 
        These two inequalities rely on an arbitrarily large constant $M$.
        Inequality \eqref{eq:lp_line_flow} forces line power flows to remain below thermal limits.
        Constraint \eqref{eq:lp_address_balance} imposes power balance.
        Finally, constraint \eqref{eq:lp_coeff} defines the coefficient $\mu$ as an affine function of $\lambda$, while constraint \eqref{eq:lp_address_slack} defines the phase origin.

                \begin{figure}[t!]
            \centering
            \begin{subfigure}[b]{0.2\linewidth}
                \centering
                \resizebox{\linewidth}{!}{\input{figures/example.tikz}}
                \caption{Base case.}
                \label{fig:base_case}
            \end{subfigure}
            \hfill
            \begin{subfigure}[b]{0.3\linewidth}
                \centering
                \resizebox{\linewidth}{!}{\input{figures/example_y1.tikz}}
                \caption{$y=[1, 1, 1, 1]$.}
                \label{fig:y1}
            \end{subfigure}    
            \hfill
            \begin{subfigure}[b]{0.3\linewidth}
                \centering
                \resizebox{\linewidth}{!}{\input{figures/example_y0.tikz}}
                \caption{$y=[1, 0, 1, 1]$.}
                \label{fig:y0}
            \end{subfigure}    
        
            \caption{Impact of the decision variable $y$ over the topology of a toy example. For a simple base case (a) with \textit{Type 1} substations, $y=[1, 1, 1, 1]$ results in a four buses system (b), while $y=[1, 0, 1, 1]$ corresponds to a five buses topology (c).}
            \label{fig:decision_variable}
        \end{figure}

    We aim at solving the following Mixed Integer Linear Program (MILP),
    \begin{align}
        y^\star(x) \in \underset{y \in \{0, 1\}^{\mathcal{E}_x^\text{switch}}}{\arg\min}\ f(y;x).
        \label{eq:initial_optimization_problem}
    \end{align}

    \subsection{Continuous Surrogate Optimization Problem}
    For a given context $x$, the MILP \eqref{eq:initial_optimization_problem} involves a binary (vector) decision variable that makes the objective function intrinsically non-differentiable.
    Recognizing this is as a case of single-step RL \cite{sutton}, we transform it as follows.
    
    First, let us introduce a surrogate decision variable $z\in \mathbb{R}^{\mathcal{E}^\text{switch}_x}$ that associates each switch in $x$ with a real-valued score.
    Let us define the conditional probability distribution $\rho(\cdot|z)$,
    \begin{align}
        \forall y\in \{0, 1\}^{\mathcal{E}^\text{switch}_x}, \quad \rho(y|z) = \prod_{e \in \mathcal{E}^\text{switch}_x} \frac{e^{y_{e} z_{e}}}{ 1 + e^{z_{e}}}. \label{eq:rho}
    \end{align}
    It associates each switch $e\in \mathcal{E}^\text{switch}_x$ with a probability $\frac{e^{z_e}}{1 + e^{z_e}}$ of being 1 (\textit{i.e.} \textit{closed}), and $\frac{1}{1 + e^{z_e}}$ of being 0 (\textit{i.e.} \textit{open}).

    Let us define $q_\beta(\cdot|x)$ as the Boltzmann distribution derived from $f$ and a temperature parameter $\beta > 0$,
    \begin{align}
        \forall y\in \{0, 1\}^{\mathcal{E}^\text{switch}_x}, \quad q_\beta(y|x) = \frac{e^{-\beta f(y;x)}}{Z_\beta(x)},
    \end{align}
    where $Z_\beta(x) = \sum_{y'} e^{-\beta f(y';x)}$.
    It associates higher probabilities to decisions $y$ that have a lower score $f(y;x)$.

    Then, let us define the surrogate objective function $f^\beta_\rho$ as the Kullback-Leibler divergence from $\rho(\cdot|z)$ to $q_\beta(\cdot|x)$,
    \begin{align}
        f^\beta_\rho(z;x) &:= D_{KL}[\rho(\cdot|z) \Vert q_\beta(\cdot|x)], \nonumber\\
            &= -H_\rho(z) + \beta \mathbb{E}_{y \sim \rho(\cdot|z)} \left[ f(y;x) \right] + \log Z_\beta(x), \label{eq:surrogate_objective}
    \end{align}
    where $H_\rho(z)$ is the entropy of $\rho(\cdot|z)$.

\begin{figure}[h!]
        \centering
        \begin{subfigure}[b]{0.32\linewidth}
            \centering
            \includegraphics[width = \linewidth]{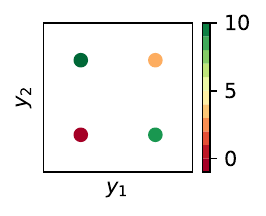}
            \caption{$y \mapsto f(y;x)$.\\ $\left. \right.$}
            \label{fig:f}
        \end{subfigure}
        \hfill
        \begin{subfigure}[b]{0.32\linewidth}
            \centering
            \includegraphics[width = \linewidth]{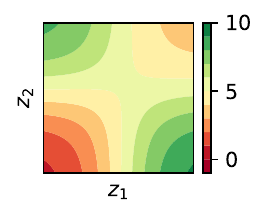}
            \caption{$z \mapsto f_\rho^\beta(z;x)/\beta$ when $\beta=1$.}
            \label{fig:frho_beta1}
        \end{subfigure}    
        \hfill
        \begin{subfigure}[b]{0.32\linewidth}
            \centering
            \includegraphics[width = \linewidth]{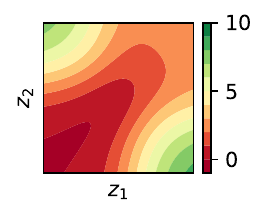}
            \caption{$z \mapsto f_\rho^\beta(z;x)/\beta$ when $\beta=0.1$.}
            \label{fig:frho_beta01}
        \end{subfigure}    
    
        \caption{Conversion of a two-dimensional combinatorial optimization problem into a continuous surrogate optimization problem.
        For a given context $x$, the decision variable $y = (y_1, y_2)$ is defined over a discrete set $\{0, 1\}^2$, as illustrated in (a).
        First, we introduce a continuous surrogate decision variable $z=(z_1, z_2)$ defined over $\mathbb{R}^2$.
        Then, we replace the initial objective function $f$ with a continuous surrogate objective function $f_\rho^\beta$ defined in equation \eqref{eq:surrogate_objective} and illustrated in (b) and (c).
        }
        \label{fig:surrogate_continuous_optimization}
    \end{figure}

    We propose to replace the initial MILP \eqref{eq:initial_optimization_problem} with the following continuous surrogate optimization problem,
    \begin{align}
        z_\beta^\star(x) \in \underset{z\in \mathbb{R}^{\mathcal{E}_x^\text{switch}}}{\arg\min}\ f^\beta_\rho(z;x).
        \label{eq:surrogate_optimization_problem}
    \end{align}
    This change of problem is motivated by the intuition that if $z$ is a good solution of problem \eqref{eq:surrogate_optimization_problem}, then $y^\star_\rho(z) \in \arg\max_y \rho(y|z)$ should also be a good solution of problem \eqref{eq:initial_optimization_problem} (see Figure \ref{fig:surrogate_continuous_optimization}). Moreover, given the factorization of eqn. \eqref{eq:rho}, the latter maximization can simply be done switch by switch. 
    {Notice that the surrogate optimization problem \eqref{eq:surrogate_optimization_problem} is not linear, and that to estimate $f^\beta_\rho$, one would need to know the values of $f$ for all possible values of $y$, which is computationally intractable.
    Hence, we propose to resort to an approximated Monte-Carlo based local resolution.}

    $f_\rho^\beta$ is continuous and differentiable \textit{w.r.t.} $z$, which enables the use of gradient-based methods.
    Using the log-trick ($\nabla_z \phi(z)=\phi(z)\nabla_z\log\phi(z)$), its derivative is expressed as:
    \begin{align}
        &g^\beta_\rho(z;x) := \nabla_z f^\beta_\rho(z;x), \nonumber \\
        &\quad = -\nabla_z H_\rho(z) + \beta \mathbb{E}_{y \sim \rho(\cdot|z)} \left[ f(y;x) \nabla_z \log \rho(y|z) \right], \nonumber \\
        &\quad = \sigma(z) \odot \sigma(\text{-}z) \odot z + \beta \mathbb{E}_{y \sim \rho(\cdot|z)} \left[ f(y;x) \left( y \text{-} \sigma(z) \right) \right]. \label{eq:surrogate_gradient}
    \end{align}
    where $\sigma(z) = \frac{e^z}{1+e^z}$ (component-wise) and $\odot$ is the Hadamard product.
    While the entropy term has a closed form expression, the second term must, in general, be estimated by the Monte Carlo (MC) method,
    \begin{align}
        \hat{g}^\text{MC} = \sigma(z) \odot \sigma(\text{-}z) \odot z + \frac{\beta}{N} \sum_{i=1}^N f_i \times (y_i - \sigma(z)), \label{eq:vanilla_gradient}
    \end{align}
    where $f_i = f(y_i;x)$, with $(y_i)_{i=1}^N$ i.i.d. samples from $\rho(\cdot|z)$.

    \subsection{Amortized Optimization}
    We aim at solving the continuous surrogate optimization problem \eqref{eq:surrogate_optimization_problem} not only for a single context $x$, but for a whole distribution $p$ of contexts.
    Thus, we introduce a mapping $\hat{z}_\theta$ -- parameterized by a vector $\theta \in \Theta$ -- that maps contexts $x$ to surrogate decisions $z$.

    We use the previously introduced H2MGNODE architecture \cite{donon2022}, a type of GNN designed to process H2MGs through a \emph{Neural Ordinary Differential Equation} (NODE) \cite{ChenRBD18}.
    It essentially relies on the continuous propagation of information between direct neighbors, by associating each address $a\in \mathcal{A}_x$ with a time-varying latent vector $h^t_a \in \mathbb{R}^d$, where $t \in \left[0, 1\right]$ is an artificial time variable and $d \in \mathbb{N}$ is the latent space dimension.
    It is defined by the following equations,
    \begin{align}
        &\!\!\!\forall c \in \mathcal{C}, \forall e\in \mathcal{E}^c_x, \Tilde{x}_e = E_\theta^c(x_e), \label{eq:h2mgnode_encoder}\\
        &\!\!\!\forall a \in \mathcal{A}_x,\! \begin{cases}
            h_a^{t=0} = [0, \dots, 0],  \\
            \! \dfrac{dh_a^t}{dt}\! =\! F_\theta \! \! \left[ h_a^t, \! \tanh\!\left(\sum\limits_{\substack{(c,e,o)\\ \in \mathcal{N}_x(a)}} \! \! M_\theta^{c,o}(h^t_e, \Tilde{x}_e) \!\! \right) \! \! \right]\!\!, \! \!
        \end{cases} \label{eq:h2mgnode_node}\\
        &\!\!\!\forall c \in \mathcal{C}, \forall e\in \mathcal{E}^c_x, [\hat{z}_\theta(x)]_e = D_\theta^c(\Tilde{x}_e, h_e^{t=1}), \label{eq:h2mgnode_decoder}
    \end{align}
    where $\mathcal{N}_x(a) = \{(c,e,o) | e\in \mathcal{E}^c_x, o(e)=a\}$ is the neighborhood of address $a$, $h_e=(h_{o(e)})_{o \in \mathcal{O}^c}$ is the concatenation of the latent vectors of addresses it is connected to and functions $(E_\theta^c)_{c\in \mathcal{C}}$, $F_\theta$, $(M^{c,o}_\theta)_{c\in \mathcal{C}, o \in \mathcal{O}^c}$ and $(D^c_\theta)_{c\in \mathcal{C}}$ are basic Multi-Layer Perceptrons (MLPs).
    
    We now consider the following \emph{Amortized Optimization} problem,
    \begin{align}
        \theta^\star \in \underset{\theta \in \Theta}{\operatorname{argmin}}\ \mathbb{E}_{x\sim p}\left[ f^\beta_\rho(\hat{z}_\theta(x);x)\right].\label{eq:amortized_problem}
    \end{align}
    For a given context $x$, the following holds,
    \begin{align}
        \nabla_\theta f^\beta_\rho(\hat{z}_\theta(x); x) = J_\theta \left[ \hat{z}_\theta \right](x)^\top . g^\beta_\rho(\hat{z}_\theta(x); x),
    \end{align}
    where $J_\theta \left[ \hat{z}_\theta \right](x)$ is the Jacobian matrix of the GNN $\hat{z}_\theta$ estimated by automatic differentiation.
    From this basic chain-rule, we derive the training loop detailed in Algorithm \ref{alg:training_loop} in the simplying case of minibatches of size 1.
    In practice, a minibatch of contexts is sampled in step (a), steps (b) and (c) are performed in parallel for each context, and the average backpropagated gradient is used in step (d).
    \begin{algorithm}
        \caption{Amortized Optimization Training Loop}\label{alg:training_loop}
        \begin{algorithmic}
        \Require $p$, $\theta$, $\hat{z}_\theta$, $\hat{g}^\text{Method}$, $\alpha$
        \While{not converged}
            \State $x \sim p$ \Comment{Context sampling (a)}
            \State $\hat{z} \gets \hat{z}_\theta(x)$ \Comment{GNN prediction (b)}
            \State $\hat{g} \gets \hat{g}^\text{Method}(\hat{z};x)$ \Comment{Gradient estimation (c)}
            \State $\theta \gets \theta - \alpha J_\theta[\hat{z}_\theta](x)^\top . \hat{g}$ \Comment{Back-propagation (d)}
        \EndWhile
        \end{algorithmic}
    \end{algorithm}

    \subsection{Surrogate Gradient Estimation}
    Unfortunately, early experiments using the gradient estimator $\hat{g}^\text{MC}$ proved unsuccessful.
    This rather simple estimator systematically gets $\theta$ stuck in a local minimum, which results in the GNN $\hat{z}_\theta$ always returning an all-closed topology.
    In what follows, we introduce two successful methods for gradient estimation (\textit{i.e.} step (c) of Algorithm \ref{alg:training_loop}).

        \paragraph{Filtered MC}
        A first idea is to filter out decisions that result in excessively large scores, to prevent them from having too much impact over the gradient estimator.
        \begin{align}
            \hat{g}^\text{FMC} = z \odot \sigma(z) \odot \sigma(\text{-}z) + \frac{\beta}{N} \sum_{i=1}^N \Tilde{f}_i \times (y_i - \sigma(z)), \label{eq:filtered_monte_carlo}
        \end{align}
        where $\Tilde{f}_i = -\sigma\left[-(f_i - \min_j f_j)/\tau\right]$ and $\tau > 0$ is a hyper-parameter.
        Equation \eqref{eq:filtered_monte_carlo} should be interpreted as follows.
        \begin{itemize}[noitemsep,topsep=0pt]
            \item If $y_i$ returns the smallest score in the current exploration round, then $f_i=\min_j f_j$, and $\Tilde{f}_i=-1/2$.
            \item If $y_i$ returns a score $f_i \gg \min_j f_j + \tau$, then $\Tilde{f}_i \approx0$.
        \end{itemize}

        \paragraph{Memory Table}
        A second idea involves a memory table $y^\text{mem}$ that associates each context $x$ from the train set with its current best sampled decision.
        At each training step, $N$ decisions are sampled for the current context $x$.
        If one of these decisions returns a better score than $y^\text{mem}(x)$, then it overwrites it.
        Finally, we replace $f$ with $(y;x)\mapsto y^\top.(2y^\text{mem}(x)-1)$ in equation \eqref{eq:surrogate_gradient}.
        Observing that for any vector $v\in \mathbb{R}^{\mathcal{E}_x^\text{switch}}$,
        \begin{align}
            \nabla_z \left( \sum_y \rho(y|z) \times y^\top . v \right)  = \left( \sum_y \nabla_z\rho(y|z) . y^\top \right) . v
        \end{align}
        and that $\sum_y \nabla_z\rho(y|z).y^\top$ is a diagonal matrix with coefficients given by vector $\sigma(z) \odot \sigma(\text{-}z)$, we obtain the following gradient estimator,
        \begin{align}
            \hat{g}^\text{MT} = \sigma(z) \odot \sigma(\text{-}z) \odot \left[ z-\beta (2y^\text{mem}(x)-1) \right]. \label{eq:memory}
        \end{align}
        Equation \eqref{eq:memory} can also be interpreted as the gradient of the Euclidean distance between $z$ and $\beta (2 y^\text{mem}(x)-1)$, weighted by $\sigma(z) \odot \sigma(\text{-}z)$.

    \subsection{Ensemble Model}
    Experimentally, the two gradient estimators $\hat{g}^\text{FMC}$ and $\hat{g}^\text{MT}$ induce very different GNN behaviors.
    We propose to combine these two models trained independantly by using the two estimators in an ensemble method \cite{Ganaie_2022}, which basically amounts to using only the better decision of the two.
    In a given context $x$, let $\hat{y}^\text{FMC}$ be the decision made by a model trained using the \textit{Filtered MC} gradient, and $\hat{y}^\text{MT}$ be the decision made by a model trained using the \textit{Memory Table} gradient.
    We define the decision of the \textit{Ensemble} model as,
    \begin{align}
        \hat{y}^\text{Ens.} = \begin{cases}
                \hat{y}^\text{FMC} &\text{if $f(\hat{y}^\text{FMC};x) < f(\hat{y}^\text{MT};x)$} \\
                \hat{y}^\text{MT} &\text{otherwise}
        \end{cases}
    \end{align}
    Notice that this third approach requires the training of two models, to perform two GNN inferences and to compute the cost function two times.

    \subsection{Sampling Policy}
    The test case illustrated in Figure \ref{fig:12_substations_system} displays multiple symmetries that can make the sampling very inefficient.
    For instance, opening a single switch in \textit{Type 2} substation is equivalent to keeping all switches closed.
    When sampling random decisions, we consider each substation separately, and reject actions that are equivalent to doing nothing with respect to the 'all closed' configuration.

%% file: figures/example.tikz
\tikzset{every picture/.style={line width=0.75pt}} 

\begin{tikzpicture}[x=0.75pt,y=0.75pt,yscale=-1,xscale=1]

\draw  [line width=1.5]  (9.77,60.12) .. controls (9.77,54.53) and (14.3,50) .. (19.88,50) .. controls (25.47,50) and (30,54.53) .. (30,60.12) .. controls (30,65.7) and (25.47,70.23) .. (19.88,70.23) .. controls (14.3,70.23) and (9.77,65.7) .. (9.77,60.12) -- cycle ;
\draw  [line width=1.5]  (90,60.12) .. controls (90,54.53) and (94.53,50) .. (100.12,50) .. controls (105.7,50) and (110.23,54.53) .. (110.23,60.12) .. controls (110.23,65.7) and (105.7,70.23) .. (100.12,70.23) .. controls (94.53,70.23) and (90,65.7) .. (90,60.12) -- cycle ;
\draw  [line width=1.5]  (50,20.12) .. controls (50,14.53) and (54.53,10) .. (60.12,10) .. controls (65.7,10) and (70.23,14.53) .. (70.23,20.12) .. controls (70.23,25.7) and (65.7,30.23) .. (60.12,30.23) .. controls (54.53,30.23) and (50,25.7) .. (50,20.12) -- cycle ;
\draw  [line width=1.5]  (50,99.88) .. controls (50,94.3) and (54.53,89.77) .. (60.12,89.77) .. controls (65.7,89.77) and (70.23,94.3) .. (70.23,99.88) .. controls (70.23,105.47) and (65.7,110) .. (60.12,110) .. controls (54.53,110) and (50,105.47) .. (50,99.88) -- cycle ;
\draw [line width=2.25]    (23.77,51) -- (51,24) ;
\draw [line width=2.25]    (28.88,56.12) -- (56,29) ;
\draw [line width=2.25]    (63.77,91) -- (91,64) ;
\draw [line width=2.25]    (68.88,96.12) -- (96,69) ;
\draw [line width=2.25]    (91.12,56.12) -- (64,29) ;
\draw [line width=2.25]    (96.12,51.12) -- (69,24) ;
\draw [line width=2.25]    (51.12,96.12) -- (24,69) ;
\draw [line width=2.25]    (56.12,91.12) -- (29,64) ;
\draw  [draw opacity=0] (10,10) -- (110.23,10) -- (110.23,110) -- (10,110) -- cycle ;

\draw (19.88,60.12) node   [align=left] {a};
\draw (60.12,20.12) node   [align=left] {b};
\draw (60.12,99.88) node   [align=left] {c};
\draw (100.12,60.12) node   [align=left] {d};

\end{tikzpicture}

%% file: figures/example_y1.tikz
\tikzset{every picture/.style={line width=0.75pt}} 

\begin{tikzpicture}[x=0.75pt,y=0.75pt,yscale=-1,xscale=1]

\draw [color={rgb, 255:red, 0; green, 0; blue, 0 }  ,draw opacity=1 ][line width=1.5]    (75,30) -- (75,13) ;
\draw [shift={(75,10)}, rotate = 90] [color={rgb, 255:red, 0; green, 0; blue, 0 }  ,draw opacity=1 ][line width=1.5]    (14.21,-4.28) .. controls (9.04,-1.82) and (4.3,-0.39) .. (0,0) .. controls (4.3,0.39) and (9.04,1.82) .. (14.21,4.28)   ;
\draw  [color={rgb, 255:red, 0; green, 0; blue, 0 }  ,draw opacity=1 ][line width=1.5]  (87.33,16.72) .. controls (88.42,15) and (89.46,13.36) .. (90.67,13.36) .. controls (91.87,13.36) and (92.91,15) .. (94,16.72) .. controls (95.09,18.44) and (96.13,20.08) .. (97.33,20.08) .. controls (98.54,20.08) and (99.58,18.44) .. (100.67,16.72) .. controls (100.67,16.72) and (100.67,16.72) .. (100.67,16.72) ;
\draw  [color={rgb, 255:red, 0; green, 0; blue, 0 }  ,draw opacity=1 ][line width=1.5]  (87.33,16.72) .. controls (87.33,13.01) and (90.32,10) .. (94,10) .. controls (97.68,10) and (100.67,13.01) .. (100.67,16.72) .. controls (100.67,20.43) and (97.68,23.44) .. (94,23.44) .. controls (90.32,23.44) and (87.33,20.43) .. (87.33,16.72) -- cycle ;

\draw [color={rgb, 255:red, 0; green, 0; blue, 0 }  ,draw opacity=1 ][line width=1.5]    (94,23.44) -- (94,30.5) ;
\draw [color={rgb, 255:red, 0; green, 0; blue, 0 }  ,draw opacity=1 ][line width=3]    (65,30) -- (105,30) ;
\draw [color={rgb, 255:red, 0; green, 0; blue, 0 }  ,draw opacity=1 ][line width=1.5]    (70,30) -- (70,35) -- (40,50) -- (40,60) ;
\draw [color={rgb, 255:red, 0; green, 0; blue, 0 }  ,draw opacity=1 ][line width=1.5]    (80,30) -- (80,40) -- (50,55) -- (50,60) ;
\draw [color={rgb, 255:red, 0; green, 0; blue, 0 }  ,draw opacity=1 ][line width=1.5]    (100,30) -- (100,35) -- (130,50) -- (130,60) ;
\draw [color={rgb, 255:red, 0; green, 0; blue, 0 }  ,draw opacity=1 ][line width=3]    (20,60) -- (55,60) ;
\draw [color={rgb, 255:red, 0; green, 0; blue, 0 }  ,draw opacity=1 ][line width=1.5]    (25,60) -- (25,77) ;
\draw [shift={(25,80)}, rotate = 270] [color={rgb, 255:red, 0; green, 0; blue, 0 }  ,draw opacity=1 ][line width=1.5]    (14.21,-4.28) .. controls (9.04,-1.82) and (4.3,-0.39) .. (0,0) .. controls (4.3,0.39) and (9.04,1.82) .. (14.21,4.28)   ;
\draw [color={rgb, 255:red, 0; green, 0; blue, 0 }  ,draw opacity=1 ][line width=3]    (115,60) -- (150,60) ;
\draw [color={rgb, 255:red, 0; green, 0; blue, 0 }  ,draw opacity=1 ][line width=1.5]    (145,60) -- (145,77) ;
\draw [shift={(145,80)}, rotate = 270] [color={rgb, 255:red, 0; green, 0; blue, 0 }  ,draw opacity=1 ][line width=1.5]    (14.21,-4.28) .. controls (9.04,-1.82) and (4.3,-0.39) .. (0,0) .. controls (4.3,0.39) and (9.04,1.82) .. (14.21,4.28)   ;
\draw [color={rgb, 255:red, 0; green, 0; blue, 0 }  ,draw opacity=1 ][line width=1.5]    (120,60) -- (120,65) -- (90,80) -- (90,90) ;
\draw [color={rgb, 255:red, 0; green, 0; blue, 0 }  ,draw opacity=1 ][line width=1.5]    (130,60) -- (130,70) -- (100,85) -- (100,90) ;
\draw [color={rgb, 255:red, 0; green, 0; blue, 0 }  ,draw opacity=1 ][line width=1.5]    (40,60) -- (40,70) -- (70,85) -- (70,90) ;
\draw [color={rgb, 255:red, 0; green, 0; blue, 0 }  ,draw opacity=1 ][line width=1.5]    (50,60) -- (50,65) -- (80,80) -- (80,90) ;
\draw [color={rgb, 255:red, 0; green, 0; blue, 0 }  ,draw opacity=1 ][line width=3]    (65,90) -- (105,90) ;
\draw [color={rgb, 255:red, 0; green, 0; blue, 0 }  ,draw opacity=1 ][line width=1.5]    (95,90) -- (95,107) ;
\draw [shift={(95,110)}, rotate = 270] [color={rgb, 255:red, 0; green, 0; blue, 0 }  ,draw opacity=1 ][line width=1.5]    (14.21,-4.28) .. controls (9.04,-1.82) and (4.3,-0.39) .. (0,0) .. controls (4.3,0.39) and (9.04,1.82) .. (14.21,4.28)   ;
\draw  [color={rgb, 255:red, 0; green, 0; blue, 0 }  ,draw opacity=1 ][line width=1.5]  (138.33,46.22) .. controls (139.42,44.5) and (140.46,42.86) .. (141.67,42.86) .. controls (142.87,42.86) and (143.91,44.5) .. (145,46.22) .. controls (146.09,47.94) and (147.13,49.58) .. (148.33,49.58) .. controls (149.54,49.58) and (150.58,47.94) .. (151.67,46.22) .. controls (151.67,46.22) and (151.67,46.22) .. (151.67,46.22) ;
\draw  [color={rgb, 255:red, 0; green, 0; blue, 0 }  ,draw opacity=1 ][line width=1.5]  (138.33,46.22) .. controls (138.33,42.51) and (141.32,39.5) .. (145,39.5) .. controls (148.68,39.5) and (151.67,42.51) .. (151.67,46.22) .. controls (151.67,49.93) and (148.68,52.94) .. (145,52.94) .. controls (141.32,52.94) and (138.33,49.93) .. (138.33,46.22) -- cycle ;

\draw [color={rgb, 255:red, 0; green, 0; blue, 0 }  ,draw opacity=1 ][line width=1.5]    (145,52.94) -- (145,60) ;
\draw  [color={rgb, 255:red, 0; green, 0; blue, 0 }  ,draw opacity=1 ][line width=1.5]  (18.33,46.72) .. controls (19.42,45) and (20.46,43.36) .. (21.67,43.36) .. controls (22.87,43.36) and (23.91,45) .. (25,46.72) .. controls (26.09,48.44) and (27.13,50.08) .. (28.33,50.08) .. controls (29.54,50.08) and (30.58,48.44) .. (31.67,46.72) ;
\draw  [color={rgb, 255:red, 0; green, 0; blue, 0 }  ,draw opacity=1 ][line width=1.5]  (18.33,46.72) .. controls (18.33,43.01) and (21.32,40) .. (25,40) .. controls (28.68,40) and (31.67,43.01) .. (31.67,46.72) .. controls (31.67,50.43) and (28.68,53.44) .. (25,53.44) .. controls (21.32,53.44) and (18.33,50.43) .. (18.33,46.72) -- cycle ;

\draw [color={rgb, 255:red, 0; green, 0; blue, 0 }  ,draw opacity=1 ][line width=1.5]    (25,53.44) -- (25,60.5) ;
\draw  [color={rgb, 255:red, 0; green, 0; blue, 0 }  ,draw opacity=1 ][line width=1.5]  (81.67,103.28) .. controls (80.58,105) and (79.54,106.64) .. (78.33,106.64) .. controls (77.13,106.64) and (76.09,105) .. (75,103.28) .. controls (73.91,101.56) and (72.87,99.92) .. (71.67,99.92) .. controls (70.46,99.92) and (69.42,101.56) .. (68.33,103.28) ;
\draw  [color={rgb, 255:red, 0; green, 0; blue, 0 }  ,draw opacity=1 ][line width=1.5]  (81.67,103.28) .. controls (81.67,106.99) and (78.68,110) .. (75,110) .. controls (71.32,110) and (68.33,106.99) .. (68.33,103.28) .. controls (68.33,99.57) and (71.32,96.56) .. (75,96.56) .. controls (78.68,96.56) and (81.67,99.57) .. (81.67,103.28) -- cycle ;

\draw [color={rgb, 255:red, 0; green, 0; blue, 0 }  ,draw opacity=1 ][line width=1.5]    (75,96.56) -- (75,89.5) ;
\draw [color={rgb, 255:red, 0; green, 0; blue, 0 }  ,draw opacity=1 ][line width=1.5]    (120,60) -- (120,55) -- (90,40) -- (90,30) ;
\draw  [draw opacity=0] (10,10) -- (160,10) -- (160,110) -- (10,110) -- cycle ;

\end{tikzpicture}

%% file: figures/example_y0.tikz
\tikzset{every picture/.style={line width=0.75pt}} 

\begin{tikzpicture}[x=0.75pt,y=0.75pt,yscale=-1,xscale=1]

\draw [color={rgb, 255:red, 0; green, 0; blue, 0 }  ,draw opacity=1 ][line width=1.5]    (75,30) -- (75,13) ;
\draw [shift={(75,10)}, rotate = 90] [color={rgb, 255:red, 0; green, 0; blue, 0 }  ,draw opacity=1 ][line width=1.5]    (14.21,-4.28) .. controls (9.04,-1.82) and (4.3,-0.39) .. (0,0) .. controls (4.3,0.39) and (9.04,1.82) .. (14.21,4.28)   ;
\draw  [color={rgb, 255:red, 0; green, 0; blue, 0 }  ,draw opacity=1 ][line width=1.5]  (87.33,16.72) .. controls (88.42,15) and (89.46,13.36) .. (90.67,13.36) .. controls (91.87,13.36) and (92.91,15) .. (94,16.72) .. controls (95.09,18.44) and (96.13,20.08) .. (97.33,20.08) .. controls (98.54,20.08) and (99.58,18.44) .. (100.67,16.72) ;
\draw  [color={rgb, 255:red, 0; green, 0; blue, 0 }  ,draw opacity=1 ][line width=1.5]  (87.33,16.72) .. controls (87.33,13.01) and (90.32,10) .. (94,10) .. controls (97.68,10) and (100.67,13.01) .. (100.67,16.72) .. controls (100.67,20.43) and (97.68,23.44) .. (94,23.44) .. controls (90.32,23.44) and (87.33,20.43) .. (87.33,16.72) -- cycle ;

\draw [color={rgb, 255:red, 0; green, 0; blue, 0 }  ,draw opacity=1 ][line width=1.5]    (94,23.44) -- (94,30.5) ;
\draw [color={rgb, 255:red, 0; green, 0; blue, 0 }  ,draw opacity=1 ][line width=3]    (65,30) -- (105,30) ;
\draw [color={rgb, 255:red, 0; green, 0; blue, 0 }  ,draw opacity=1 ][line width=1.5]    (70,30) -- (70,35) -- (40,50) -- (40,60) ;
\draw [color={rgb, 255:red, 0; green, 0; blue, 0 }  ,draw opacity=1 ][line width=1.5]    (70,60) -- (70,55) -- (50,55) -- (50,60) ;
\draw [color={rgb, 255:red, 0; green, 0; blue, 0 }  ,draw opacity=1 ][line width=1.5]    (100,30) -- (100,35) -- (130,50) -- (130,60) ;
\draw [color={rgb, 255:red, 0; green, 0; blue, 0 }  ,draw opacity=1 ][line width=3]    (20,60) -- (55,60) ;
\draw [color={rgb, 255:red, 0; green, 0; blue, 0 }  ,draw opacity=1 ][line width=1.5]    (25,60) -- (25,77) ;
\draw [shift={(25,80)}, rotate = 270] [color={rgb, 255:red, 0; green, 0; blue, 0 }  ,draw opacity=1 ][line width=1.5]    (14.21,-4.28) .. controls (9.04,-1.82) and (4.3,-0.39) .. (0,0) .. controls (4.3,0.39) and (9.04,1.82) .. (14.21,4.28)   ;
\draw [color={rgb, 255:red, 0; green, 0; blue, 0 }  ,draw opacity=1 ][line width=3]    (115,60) -- (150,60) ;
\draw [color={rgb, 255:red, 0; green, 0; blue, 0 }  ,draw opacity=1 ][line width=1.5]    (145,60) -- (145,77) ;
\draw [shift={(145,80)}, rotate = 270] [color={rgb, 255:red, 0; green, 0; blue, 0 }  ,draw opacity=1 ][line width=1.5]    (14.21,-4.28) .. controls (9.04,-1.82) and (4.3,-0.39) .. (0,0) .. controls (4.3,0.39) and (9.04,1.82) .. (14.21,4.28)   ;
\draw [color={rgb, 255:red, 0; green, 0; blue, 0 }  ,draw opacity=1 ][line width=1.5]    (120,60) -- (120,65) -- (90,80) -- (90,90) ;
\draw [color={rgb, 255:red, 0; green, 0; blue, 0 }  ,draw opacity=1 ][line width=1.5]    (130,60) -- (130,70) -- (100,85) -- (100,90) ;
\draw [color={rgb, 255:red, 0; green, 0; blue, 0 }  ,draw opacity=1 ][line width=1.5]    (40,60) -- (40,70) -- (70,85) -- (70,90) ;
\draw [color={rgb, 255:red, 0; green, 0; blue, 0 }  ,draw opacity=1 ][line width=1.5]    (50,60) -- (50,65) -- (80,80) -- (80,90) ;
\draw [color={rgb, 255:red, 0; green, 0; blue, 0 }  ,draw opacity=1 ][line width=3]    (65,90) -- (105,90) ;
\draw [color={rgb, 255:red, 0; green, 0; blue, 0 }  ,draw opacity=1 ][line width=1.5]    (95,90) -- (95,107) ;
\draw [shift={(95,110)}, rotate = 270] [color={rgb, 255:red, 0; green, 0; blue, 0 }  ,draw opacity=1 ][line width=1.5]    (14.21,-4.28) .. controls (9.04,-1.82) and (4.3,-0.39) .. (0,0) .. controls (4.3,0.39) and (9.04,1.82) .. (14.21,4.28)   ;
\draw  [color={rgb, 255:red, 0; green, 0; blue, 0 }  ,draw opacity=1 ][line width=1.5]  (138.33,46.22) .. controls (139.42,44.5) and (140.46,42.86) .. (141.67,42.86) .. controls (142.87,42.86) and (143.91,44.5) .. (145,46.22) .. controls (146.09,47.94) and (147.13,49.58) .. (148.33,49.58) .. controls (149.54,49.58) and (150.58,47.94) .. (151.67,46.22) .. controls (151.67,46.22) and (151.67,46.22) .. (151.67,46.22) ;
\draw  [color={rgb, 255:red, 0; green, 0; blue, 0 }  ,draw opacity=1 ][line width=1.5]  (138.33,46.22) .. controls (138.33,42.51) and (141.32,39.5) .. (145,39.5) .. controls (148.68,39.5) and (151.67,42.51) .. (151.67,46.22) .. controls (151.67,49.93) and (148.68,52.94) .. (145,52.94) .. controls (141.32,52.94) and (138.33,49.93) .. (138.33,46.22) -- cycle ;

\draw [color={rgb, 255:red, 0; green, 0; blue, 0 }  ,draw opacity=1 ][line width=1.5]    (145,52.94) -- (145,60) ;
\draw  [color={rgb, 255:red, 0; green, 0; blue, 0 }  ,draw opacity=1 ][line width=1.5]  (18.33,46.72) .. controls (19.42,45) and (20.46,43.36) .. (21.67,43.36) .. controls (22.87,43.36) and (23.91,45) .. (25,46.72) .. controls (26.09,48.44) and (27.13,50.08) .. (28.33,50.08) .. controls (29.54,50.08) and (30.58,48.44) .. (31.67,46.72) ;
\draw  [color={rgb, 255:red, 0; green, 0; blue, 0 }  ,draw opacity=1 ][line width=1.5]  (18.33,46.72) .. controls (18.33,43.01) and (21.32,40) .. (25,40) .. controls (28.68,40) and (31.67,43.01) .. (31.67,46.72) .. controls (31.67,50.43) and (28.68,53.44) .. (25,53.44) .. controls (21.32,53.44) and (18.33,50.43) .. (18.33,46.72) -- cycle ;

\draw [color={rgb, 255:red, 0; green, 0; blue, 0 }  ,draw opacity=1 ][line width=1.5]    (25,53.44) -- (25,60.5) ;
\draw  [color={rgb, 255:red, 0; green, 0; blue, 0 }  ,draw opacity=1 ][line width=1.5]  (81.67,103.28) .. controls (80.58,105) and (79.54,106.64) .. (78.33,106.64) .. controls (77.13,106.64) and (76.09,105) .. (75,103.28) .. controls (73.91,101.56) and (72.87,99.92) .. (71.67,99.92) .. controls (70.46,99.92) and (69.42,101.56) .. (68.33,103.28) ;
\draw  [color={rgb, 255:red, 0; green, 0; blue, 0 }  ,draw opacity=1 ][line width=1.5]  (81.67,103.28) .. controls (81.67,106.99) and (78.68,110) .. (75,110) .. controls (71.32,110) and (68.33,106.99) .. (68.33,103.28) .. controls (68.33,99.57) and (71.32,96.56) .. (75,96.56) .. controls (78.68,96.56) and (81.67,99.57) .. (81.67,103.28) -- cycle ;

\draw [color={rgb, 255:red, 0; green, 0; blue, 0 }  ,draw opacity=1 ][line width=1.5]    (75,96.56) -- (75,89.5) ;
\draw [color={rgb, 255:red, 0; green, 0; blue, 0 }  ,draw opacity=1 ][line width=1.5]    (120,60) -- (120,55) -- (100,55) -- (100,60) ;
\draw [color={rgb, 255:red, 0; green, 0; blue, 0 }  ,draw opacity=1 ][line width=1.5]    (77,60) -- (77,43) ;
\draw [shift={(77,40)}, rotate = 90] [color={rgb, 255:red, 0; green, 0; blue, 0 }  ,draw opacity=1 ][line width=1.5]    (14.21,-4.28) .. controls (9.04,-1.82) and (4.3,-0.39) .. (0,0) .. controls (4.3,0.39) and (9.04,1.82) .. (14.21,4.28)   ;
\draw  [color={rgb, 255:red, 0; green, 0; blue, 0 }  ,draw opacity=1 ][line width=1.5]  (85.33,46.72) .. controls (86.42,45) and (87.46,43.36) .. (88.67,43.36) .. controls (89.87,43.36) and (90.91,45) .. (92,46.72) .. controls (93.09,48.44) and (94.13,50.08) .. (95.33,50.08) .. controls (96.54,50.08) and (97.58,48.44) .. (98.67,46.72) .. controls (98.67,46.72) and (98.67,46.72) .. (98.67,46.72) ;
\draw  [color={rgb, 255:red, 0; green, 0; blue, 0 }  ,draw opacity=1 ][line width=1.5]  (85.33,46.72) .. controls (85.33,43.01) and (88.32,40) .. (92,40) .. controls (95.68,40) and (98.67,43.01) .. (98.67,46.72) .. controls (98.67,50.43) and (95.68,53.44) .. (92,53.44) .. controls (88.32,53.44) and (85.33,50.43) .. (85.33,46.72) -- cycle ;

\draw [color={rgb, 255:red, 0; green, 0; blue, 0 }  ,draw opacity=1 ][line width=1.5]    (92,53.44) -- (92,60.5) ;
\draw [color={rgb, 255:red, 0; green, 0; blue, 0 }  ,draw opacity=1 ][line width=3]    (65,60) -- (105,60) ;
\draw  [draw opacity=0] (10,10) -- (160,10) -- (160,110) -- (10,110) -- cycle ;

\end{tikzpicture}

%% file: 3_case_study.tex
\section{Case Study}
\label{sec:case_study}

Let us now experimentally validate our methodology on the 12 substations test case from Figure \ref{fig:patricks_case}.
Substations b, c and j are of \textit{Type 1}, while substations a, d, e, f, g, h, i, k and l are of \textit{Type 2}.

    \subsection{Datasets}

        In order to train and compare our models against a varied distribution $p$ of contexts $x$, a large dataset should first be generated.
        {The following sampling methodology and the choice of its parameters ensure a sufficient variability in the solutions found by a classical optimization baseline.}
        We start from an initial operating condition defined in \cite{patrick} and illustrated in Figure \ref{fig:patricks_case} and add a random noise over injections and thermal limits and randomly disconnect transmission lines.
        Values from the source case are indexed by a superscript $0$.
        
        Let $e$ be a hyper-edge of class $c \in \{ \text{Load}, \text{Gen} \}$ in area $Z_k \in \{ Z_1, Z_2 \}$.
        Its active power $P_e$ is sampled as follows,
        \begin{align}
            P_e = \frac{\epsilon_e}{\sum_{e'\in \mathcal{E}^c_x }\mathbb{1}_{e'}^{Z_k} \epsilon_{e'} } \times \frac{\epsilon^c_k}{\epsilon^c_1+\epsilon^c_2} \times \epsilon_T,
        \end{align}
        where $\forall e'\in \mathcal{E}^c_x, \epsilon_{e'} \sim \mathcal{N}(P^0_{e'}, \sigma_L)$ is a local noise, $\forall k' \in \{ 1, 2 \}, \epsilon_{k'} \sim \mathcal{N}(\sum_{e' \in \mathcal{E}^c} \mathbb{1}_{e'}^{Z_{k'}} P^0_{e'}, \sigma_Z)$ is a zone-specific noise, and $\epsilon_T \sim \mathcal{N}(\sum_{e' \in \mathcal{E}^c_x} P^0_{e'}, \sigma_T)$ is a global noise.
        Standard deviations are set to $\sigma_L = 50$MW, $\sigma_Z = 200$MW and $\sigma_T = 500$MW.

        {
        In the source case, thermal limits have 3 possible values, $\overline{F}^0_1$, $\overline{F}^0_2$ and $\overline{F}^0_\text{border}$ for lines respectively in $Z_1$, $Z_2$ and at the border.
        Thermal limits are randomized so that lines in area $Z_1$ have the same thermal limit sampled from $\mathcal{N}(\overline{F}^0_1, \sigma_{F})$, lines in area $Z_2$ have the same thermal limit sampled from $\mathcal{N}(\overline{F}^0_2, \sigma_{F})$, and lines at the border have the same thermal limit sampled from $\mathcal{N}(\overline{F}^0_\text{border}, \sigma_{F})$. Standard deviation is set to $\sigma_{F}=50$MW.}

        {
        Finally, topological variability is ensured by randomly disconnecting exactly one uniformly chosen line with a 60\% probability, and exactly two uniformly chosen lines with a 10\% probability.}

        {
        Four datasets are thus generated, namely \textit{Large Train} (850k contexts), \textit{Small Train} (10k contexts), \textit{Validation} (100k contexts) and \textit{Test} (10k contexts).}
        
    \subsection{Baselines}

    Our methods \textit{Filtered MC}, \textit{Memory Table} and \textit{Ensemble}  are compared to the following baselines.

        \paragraph{All Closed}
        All switches are closed.

        \paragraph{MILP Solver}
        The initial optimization problem \eqref{eq:initial_optimization_problem} is solved using XPRESS \cite{fico_xpress}.
        In order to keep computational times reasonably low, the problem is restricted to at most 6 switch openings and a 1\%-gap convergence criterion and a 10-minutes time limit are applied.
        Parameter $M$ from equations \eqref{eq:lp_breaker_closed} and \eqref{eq:lp_breaker_open} is set to $\sum_{e\in \mathcal{E}^\text{gen}_x}P_e$.

        \paragraph{GNN Supervised}
        The third baseline consists in the imitation of the \textit{MILP Solver} in a supervised fashion (logistic regression).
        The computation of these solutions being quite expensive, the model was trained on the \textit{Small Train} set.

                    \begin{table*}[t!]\renewcommand{\tabcolsep}{1.6mm}
            \centering
            \begin{tabular}{lcccccc}
                                                    & All Closed    & MILP Solver     & GNN Supervised               & GNN Filtered MC                  & GNN Memory Table                 & GNN Ensemble               \\
                \hline
                Mean Exchange Capacity per Context  & 10.26         & 11.62           & 10.75 \tiny{$\pm$ 0.0026}    & 10.93 \tiny{$\pm$ 0.0050}        & 10.95 \tiny{$\pm$ 0.039}         & 11.14 \tiny{$\pm$ 0.014}   \\
                Mean Normalized Capacity per Context& 0.0           & 1.0             & 0.26 \tiny{$\pm$ 0.014}      & 0.26 \tiny{$\pm$ 0.0014}         & 0.41 \tiny{$\pm$ 0.0029}         & 0.50 \tiny{$\pm$ 0.021}    \\
                Mean Improvement over \textit{All Closed} & 0.00    & 15.2\%          & 5.55\% \tiny{$\pm$ 0.028}    & 8.13\% \tiny{$\pm$ 0.075}        & 7.42\% \tiny{$\pm$ 0.64}         & 10.2\% \tiny{$\pm$ 0.12}   \\
                Mean Number of Openings per Context & 0.0           & 6.0             & 2.4 \tiny{$\pm$ 0.33}        & 1.1 \tiny{$\pm$ 0.045}           & 2.1 \tiny{$\pm$ 0.20}            & 2.1 \tiny{$\pm$ 0.18}      \\
                Mean Usage Percentage per Switch    & 0.00\%        & 10.5\%         & 4.18\% \tiny{$\pm$ 0.58\%}   & 1.97\% \tiny{$\pm$ 0.079}        & 3.74\% \tiny{$\pm$ 0.34}         & 3.73\% \tiny{$\pm$ 0.31}   \\
                Number of Switches Never Used       & 57            & 0.0               & 8.33 \tiny{$\pm$ 2.36}       & 20.33 \tiny{$\pm$ 3.30}          & 12.66 \tiny{$\pm$ 0.47}          & 11 \tiny{$\pm$ 0.82} 
            \end{tabular}
            \caption{
                Score and behavior of the different models and baselines on the \textit{Test} set.
                Exchange capacities are expressed in p.u. using the $100$MW base power.
                GNN metrics are averaged over 3 distinct trainings with different random seeds.
                Standard deviation over the 3 models are reported.
                For a given context $x_i$, the normalized score is defined as $(f_i-f_i^\text{Closed}) / (f_i^\text{Solver}-f_i^\text{Closed})$, where $f_i$ is the score of the considered method and $f_i^\text{Closed}$ (resp. $f_i^\text{Solver}$) is the score of the \textit{All Closed} (resp. \textit{MILP Solver}) baseline.
                The \textit{MILP Solver} returned worse or equal performance as the \textit{All Closed} approach in 9 snapshots out of 10k, which are therefore excluded from the results analysis.
            }
            \label{tab:results}
        \end{table*}

    \subsection{Experimental Setup}

        Experiments are conducted using our {open-source package \textit{EnerGNN}}\footnote{{https://github.com/energnn/energnn}}, that relies on JAX \cite{jax2018github} and Flax \cite{flax2020github}.
        Input features are normalized using a piecewise linear approximation of the empirical cumulative distribution function, as detailed in the supplementary material of \cite{donon2024}.
        All H2MGNODE models share the same hyper-parameters, which have been adjusted by trial-and-error for lack of time to perform a comprehensive hyper-parameter optimization.
        Encoders $(E^c_\theta)_{c\in \mathcal{C}_x}$ are Multi Layer Perceptrons (MLPs) with two hidden layers of sizes $(128, 128)$, an output dimension of size $64$ and Leaky ReLU activation functions.
        Addresses latent vectors are of size $64$.
        System \eqref{eq:h2mgnode_node} is solved by Diffrax \cite{kidger2021on}, using an explicit 1$^\text{st}$-order Euler scheme with $\Delta t=0.05$, and backpropagation is performed using the recursive checkpoint adjoint method.
        Function $F_\theta$ is an MLP without any hidden layer with a Leaky ReLU applied over its output.
        Message passing functions $(M^{c,o}_\theta)_{c\in \mathcal{C}, o \in \mathcal{O}^c}$ and decoders $(D_\theta^c)_{c \in \mathcal{C}}$ are MLPs with hidden layers of sizes $(128, 128)$ and Leaky ReLU activation functions.
        Back-propagated gradients are first clipped element-wise with a $[-0.04, 0.04]$ range and then processed by Adam \cite{kingma2017} with a learning rate of $3\times 10^{-4}$ and standard parameters.

        The \textit{Filtered MC} is trained over the \textit{Large Train} set with a surrogate gradient $\hat{g}^\text{FMC}$ using $N=32$, $\tau=20$ and $\beta=0.1$.

        The \textit{Memory Table} is trained over the \textit{Small Train} set with a surrogate gradient $\hat{g}^\text{MT}$ estimated using $N=32$ and $\beta=1$. A larger dataset makes the memory table too long to populate and update, and drastically slows training down.
        Exploration is done by sampling around the current most probable decision $\arg\max_y \rho(y|z)$ and disconnecting one (resp. two) switch with probability 30\% (resp. 40\%).

        All GNNs are trained 3 times, each time with a different random seed.
        Three distinct \textit{Ensemble} models are created from the three seed pairs.
        Results in Figure \ref{fig:results} are given for one of these seeds, while results in Table \ref{tab:results} provide the mean and standard deviations over the 3 runs.

        Each training lasts between 120k and 180k iterations, with minibatches of size 8, which takes 7 to 10 days using an NVIDIA V100S GPU for forward and backward GNN computations and 8 Intel Xeon Gold 6226R CPUs @ 2.90GHz for exploration and gradient estimation.
        Models are evaluated over the \textit{Validation} set after each epoch and after every 1k iterations, and only the model with the best mean exchange capacity is kept, which can be seen as a form of early stopping.

    \subsection{Results}

        \begin{figure}[b!]
        \vspace{-2\baselineskip}

            \centering
            \begin{subfigure}[b]{\linewidth}
                \centering
                \includegraphics[width = \linewidth]{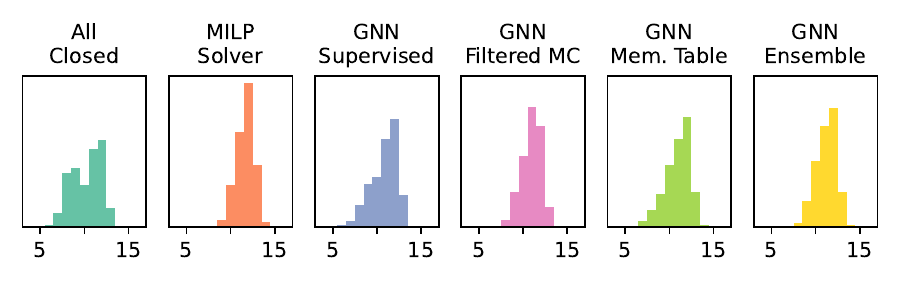}
                \vspace{-2\baselineskip}
                \caption{Histograms of exchange capacities, $-f_i=-f(y_i;x_i)$.}
                \label{fig:score}
            \end{subfigure}
            \vfill
    
            \begin{subfigure}[b]{\linewidth}
                \centering
                \includegraphics[width = \linewidth]{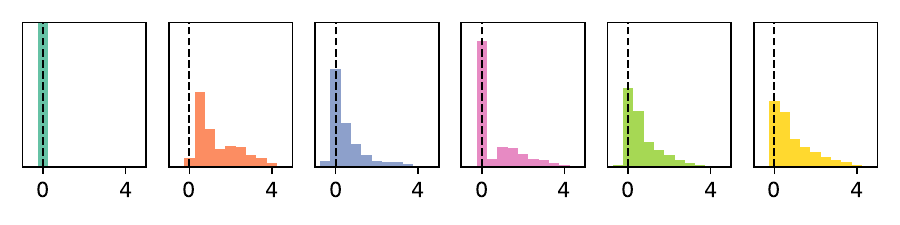}
                \vspace{-2\baselineskip}
                \caption{Histograms of capacity differences with all-closed, $f_i^\text{Closed} - f_i$.}
                \label{fig:score_minus_init}
            \end{subfigure}
            \vfill
            \begin{subfigure}[b]{\linewidth}
                \centering
                \includegraphics[width = \linewidth]{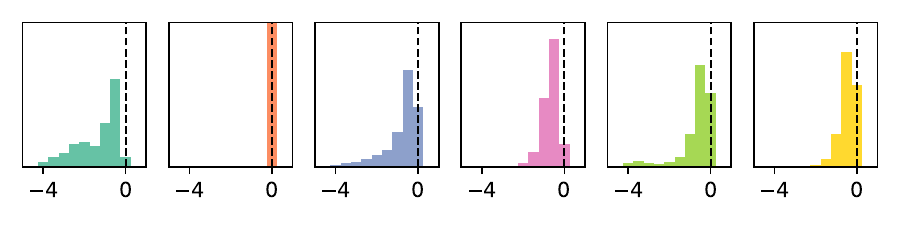}
                \vspace{-2\baselineskip}
                \caption{Histograms of capacity differences with the solver, $f_i^\text{Solver} - f_i$.}
                \label{fig:score_minus_solver}
            \end{subfigure}    
            \vfill
            \begin{subfigure}[b]{\linewidth}
                \centering
                \includegraphics[width = \linewidth]{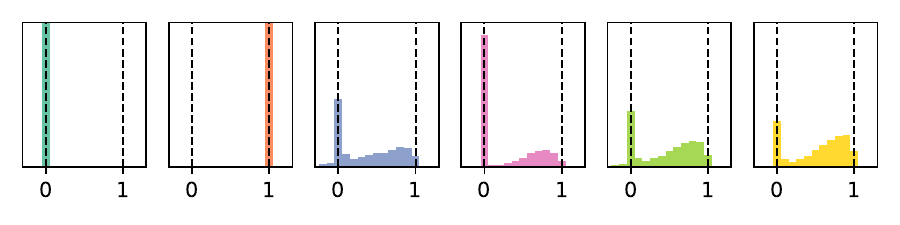}
                \vspace{-2\baselineskip}
                \caption{Histograms of normalized scores, $(f_i^\text{Closed} - f_i) / (f_i^\text{Closed} - f_i^\text{Solver})$}
                \label{fig:score_normalized}
            \end{subfigure}    
            \vfill
            \begin{subfigure}[b]{\linewidth}
                \centering
                \includegraphics[width = \linewidth]{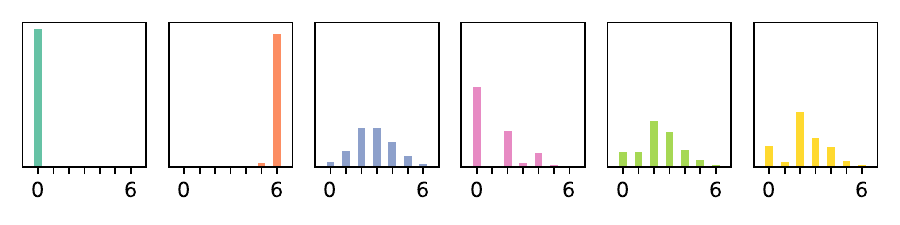}
                \vspace{-2\baselineskip}
                \caption{Histograms of the number of openings per context.}
                \label{fig:opening_count_per_context}
            \end{subfigure}    
            \vfill
            \begin{subfigure}[b]{\linewidth}
                \centering
                \includegraphics[width = \linewidth]{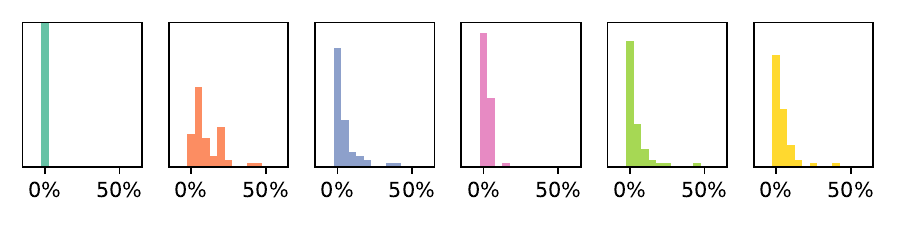}
                \vspace{-2\baselineskip}
                \caption{Histograms of opening percentage per switch.}
                \label{fig:opening_ratio_per_object}
            \end{subfigure}    
    
            \caption{
                Score and behavior histograms of the different models along with the baselines.
                Histograms (a), (b), (c), (d) et (e) are computed over the whole Test set (one value per context).
                Histogram (f) is computed over the different switches (one value per switch).
                The scale of the vertical axis is linear, and is shared between histograms of the same row.
                In (b), (c), (d) and (f), Dirac peaks are cropped for the sake of readability.
            }
            \label{fig:results}
        \end{figure}

        Figures \ref{fig:score}, \ref{fig:score_minus_init}, \ref{fig:score_minus_solver} and \ref{fig:score_normalized} display the resulting exchange capacity histograms through different perspectives.
        The \textit{Filtered MC} and \textit{Memory Table} provide comparable results, outperforming the \textit{Supervised} approach.
        As expected, the \textit{MILP Solver} is a very hard to match baseline with a 15.2\% average capacity improvement over the \textit{All Closed baseline}, although the \textit{Ensemble} model reduces the gap by performing a 10.2\% average improvement.
        In some contexts, the \textit{Memory Table} and \textit{Supervised} models degrade the exchange capacity \textit{w.r.t.} the \textit{All Closed} baseline.
        In even rarer cases (not displayed), those two models provide decisions for which there is no feasible solution to problem \eqref{eq:lp}.
        On the other hand the \textit{Filtered MC} and \textit{Ensemble} models consistently improve capacities.

        Figure \ref{fig:opening_count_per_context} underlines behavior differences between models.
        The \textit{MILP Solver} opens as many switches as possible (\textit{i.e.} 6.0), the \textit{Supervised}, \textit{Memory Table} and \textit{Ensemble} models open around 2 switches per context, while the \textit{Filtered MC} opens only approximately 1 switch on average.
        In around 50\% of contexts, the \textit{Filtered MC} prefers to do nothing, while it is extremely rare for the other models.

        Finally, Figure \ref{fig:opening_ratio_per_object} shows the distribution of usage percentage for each switch.
        The \textit{MILP Solver} uses each switch in 10.5\% of contexts on average, with some switches used in more than 40\% of contexts and 0 never-used switches.
        The \textit{Supervised}, \textit{Memory Table} and \textit{Ensemble} models only use each switch in $\sim$4\% of contexts, with $\sim$10 never-used switches.
        On the other hand, the \textit{Filtered MC} model is more frugal, using each switch in 1.97\% of contexts on average, with 20.33 never-used switches.

        In terms of computing times, all GNN inferences take $\sim$100 ms for a batch of size 8 on an NVIDIA V100S, while the \textit{MILP Solver} regularly hits the 10-minutes limit on an Intel Xeon Gold 6226R CPU @ 2.90GHz.
        Notice that the \textit{Ensemble} model requires two GNN forward passes and two objective function evaluations.

%% file: 4_conclusion.tex
\section{Conclusion}
\label{sec:conclusion}

This paper considers the problem of maximizing the exchange capacity between two neighboring areas, through optimal substation reconfiguration.
Our proposed method consists in the self-supervised training of a \textit{Graph Neural Network} to maximize exchange capacity by trial and error, falling in line with the \emph{Amortized Optimization} literature \cite{amortizedoptimization}.
Three versions of the approach are introduced and experimentally validated against an artificially-generated dataset.
All models substantially improve the mean exchange capacity compared to the \textit{All Closed} baseline, with reasonably low inference time ($\sim$100 ms).
The methodological contribution can easily be extended to other use cases such as overflow mitigation or the maximization of the exchange capacity between more than two TSOs.
This paper paves the way for leveraging modern machine learning to reach an optimality/speed compromise compatible with real-time and full-scale topology optimization of large-scale transmission systems.

Concurrently to this work, the scalability of the GNN architecture and training pipeline has been proven on the full HV-EHV French system.
Still, some adjustments are required to make the exploration and gradient estimation scalable to larger systems (\textit{cf} step (c) Algorithm \ref{alg:training_loop}).
With a larger system, the combinatorial aspect of the problem will probably prevent us from sampling good topological decisions.
One possibility would be to exploit the recent line of work on GPU-acceleration of DC simulations to explore more decisions.
We could also resort to preprocessing heuristics to reduce the decision space to subparts of the system.
Finally, the proposed methodology could be used to select substations which should then be optimized by a MILP solver.